\documentclass[twocolumn]{autart}    
\DeclareMathAlphabet{\mathpzc}{OT1}{pzc}{m}{it}
\usepackage{ifpdf}
\ifpdf
 \else
 \fi
 \usepackage{multirow}
 \usepackage[noadjust]{cite}

\usepackage{amsmath} 
\usepackage{amssymb}  
\usepackage{mathrsfs} 
\usepackage{amsfonts} 
\usepackage{cite}
\usepackage{graphicx}   
\usepackage{epstopdf}
\usepackage{xcolor}
\usepackage{subfigure}
\usepackage{algorithm}
\usepackage{etoolbox}

\let\classAND\AND
\let\AND\relax
\usepackage{algorithmic}

\let\AND\classAND
\AtBeginEnvironment{algorithmic}{\let\AND\algoAND}

\newtheorem{lemma}{Lemma}
\newtheorem{remark}{Remark}

\newtheorem{definition}{Definition}
\newtheorem{theorem}{Theorem}

\newtheorem{assumption}{Assumption}

\begin{document}

\begin{frontmatter}

\title{Secure Distributed Consensus Estimation under False Data Injection Attacks: A Defense Strategy Based on Partial Channel Coding} 
\thanks[footnoteinfo]{This work was supported by the National Natural Science Foundation of China (Grant Nos. 62503431, 62293502, 62233005, 62336005, 62573198), the National Key R\&D Program of China under grant 2023YFF1204805, the Cyprus Academy of Sciences, Letters and Arts, the Fundamental Research Funds of Zhejiang University of Science and Technology under grant 2025QN016, and the Shanghai Institute for Mathematics and Interdisciplinary Sciences (SIMIS) under grant SIMIS-ID-2025-SP (Corresponding authors: Wen Yang, Fangfei Li, and Yang Tang).}

\thanks[footnoteinfo2]{During the preparation of this manuscript, Jiahao Huang was affiliated with East China University of Science and Technology.}

\author[ZUSTx]{Jiahao Huang}\ead{Hjhao@mail.ecust.edu.cn},
\author[CYP]{Marios M. Polycarpou}\ead{mpolycar@ucy.ac.cy},
\author[ECUSTa]{Wen Yang}\ead{weny@ecust.edu.cn},
\author[ECUSTm,ECUSTa]{Fangfei Li}\ead{li$\_$fangfei@163.com},
\author[ECUSTa]{Yang Tang}\ead{yangtang@ecust.edu.cn}

\address[ZUSTx]{School of Automation and Electrical Engineering, Zhejiang University of Science and Technology, Hangzhou 310023, China}
\address[ECUSTa]{Key Laboratory of Smart Manufacturing in Energy Chemical Process, Ministry of Education, East China University of Science and Technology, Shanghai 200237, China}
\address[CYP]{KIOS Research and Innovation Center of Excellence, and Department of Electrical and Computer Engineering, University of Cyprus, Nicosia, 1678, Cyprus}
\address[ECUSTm]{School of mathematics, East China University of Science and Technology, Shanghai 200237, China}


\begin{abstract}                          
This article investigates the security issue caused by false data injection attacks in distributed estimation, wherein each sensor can construct two types of residues based on local estimates and neighbor information, respectively. The resource-constrained attacker can select partial channels from the sensor network and arbitrarily manipulate the transmitted data. We derive necessary and sufficient conditions to reveal system vulnerabilities, under which the attacker is able to diverge the estimation error while preserving the stealthiness of all residues. We propose two defense strategies with mechanisms of exploiting the Euclidean distance between local estimates to detect attacks, and adopting the coding scheme to protect the transmitted data, respectively. It is proven that the former has the capability to address the majority of security loopholes, while the latter can serve as an additional enhancement to the former. By employing the time-varying coding matrix to mitigate the risk of being cracked, we demonstrate that the latter can safeguard against adversaries injecting stealthy sequences into the encoded channels. Hence, drawing upon the security analysis, we further provide a procedure to select security-critical channels that need to be encoded, thereby achieving a trade-off between security and coding costs. Finally, some numerical simulations are conducted to demonstrate the theoretical results.
\end{abstract}

\end{frontmatter}

\section{Introduction}\label{Section1}
In Cyber-Physical Systems (CPSs), malicious third parties may compromise the cyber layer to damage the physical infrastructure, resulting in catastrophic consequences such as economic losses and casualties. Hence, the security issue is of the utmost importance to CPSs, and has attracted widespread attention from both academia and industry in recent years \cite{sandberg2015cyberphysical}. Generally, cyber attacks in CPSs can be categorized into Denial-of-Service (DoS) attack, false data injection attack, and replay attack \cite{teixeira2012attack}. Based on various attack models, the existing literature is dedicated to studying attack strategies and defense countermeasures in terms of estimation and control. The study of attack strategies helps to explore vulnerabilities in CPSs, which is also a prerequisite for designing protective measures. For energy-constrained DoS attacks, the optimal scheduling strategy to maximize the degradation of system performance was studied in \cite{zhang2015optimal,qin2020optimal}. In \cite{guo2018worst,ren2021kullback,bai2017kalman,gheitasi2022undetectable,hu2018state,mo2010false}, the authors explored stealthy strategies for false data injection attacks to evade detection, and researched the trade-off between performance degradation and attack stealthiness. The feasibility conditions of deceiving the replay attack detection was studied in \cite{mo2009secure}. Recent advances in defense methods encompass enhancing system resilience against attacks \cite{fawzi2014secure,zhu2013performance,degue2022stealthy,khazraei2022attack}, deploying novel detection mechanisms to detect attacks \cite{ahmadi2022new,li2017detection,huang2020secure,miao2016coding,griffioen2020moving}, and so on. For instance, a watermarking strategy was proposed in \cite{huang2020secure} to protect the remote state estimation from linear attacks. In \cite{miao2016coding}, the authors put forward a coding scheme to assist the $\chi^2$ detector against stealthy attacks in networked control systems. In \cite{griffioen2020moving}, a moving target defense method was proposed to break the attack stealthiness by introducing stochastic and time-varying parameters.

The above works revolve around the security of single-sensor systems or centralized sensor networks. With the advantage of openness and scalability, the distributed sensor network is also an indispensable part of CPSs. Accordingly, various types of distributed estimation algorithms have been well developed and widely used in many application areas of CPSs such as autonomous drone swarms and smart grids. For instance, an information-weighted consensus filter was proposed in \cite{kamal2013information}, which can achieve consensus of local estimates through multiple communication iterations per time instant. By taking the information pairs (matrix-vector) as the transmission data, the fusion algorithms in \cite{battistelli2014consensus} and \cite{wang2017convergence} were proposed to stabilize estimation errors under the global detectability condition. For undirected sensor networks, a distributed Kalman filter based on consensus and innovation was proposed in \cite{das2016consensus+}, and its optimal gains were also derived to minimize estimation errors. In \cite{olfati2009kalman,yang2014stochastic,yang2017stochastic}, a distributed consensus filter for directed sensor networks was proposed, requiring transmission of information vectors only once per time instant. Notice that among a variety of estimation algorithms, choosing which one is a trade-off between detectability condition, topology assumption, communication cost, and so on. For examples, the one in \cite{olfati2009kalman,yang2014stochastic,yang2017stochastic} requires a stronger detectability condition than those in \cite{kamal2013information,battistelli2014consensus,wang2017convergence,das2016consensus+}, but it is superior in saving communication costs.

However, due to the high connectivity of sensor networks, any undetected attack may spread its negative impact to the entire network. Thus, the distributed architecture is also vulnerable to cyber attacks, and its security issue has become a focal topic in the past few years \cite{mustafa2022secure,ju2020distributed,sui2021vulnerability,lu2019malicious,an2019distributed,chen2018resilient,forti2017distributed,ao2018distributed,choraria2022design,shefaei2021revealing,liu2020false,yang2019distributed,zhou2022watermarking,zhou2022security,basit2022distributed,basit2023dynamic}. For instance, a neural-network-based unified framework was introduced in \cite{basit2023dynamic} to address the distributed state and parameter estimation problem subject to deception attacks and unknown nonlinearities. Besides, in \cite{basit2022distributed}, an event-triggered distributed estimator was designed for nonlinear systems under non-periodic DoS attacks and unknown inputs. In \cite{sui2021vulnerability}, the authors studied the vulnerability of distributed estimator under stealthy attacks that can partially or fully manipulate sensor nodes. For the distributed consensus filter in \cite{olfati2009kalman,yang2014stochastic,yang2017stochastic}, the authors in \cite{liu2020false} analyzed its worst-case performance degradation under stealthy attacks that can falsify all measurements. In \cite{yang2019distributed}, a stochastic protector was proposed for the distributed consensus filter to defend against stealthy attacks, which can randomly inject Gaussian noises to the transmitted data. When each sensor adopts neighbor information to construct residues, the authors in \cite{zhou2022watermarking,zhou2022security} investigated blind spots of attack detection in the distributed consensus filter, from which stealthy attacks tampering with all channels can even destabilize the distributed estimation.

\begin{figure}[!tbp]
\begin{minipage}[t]{1\linewidth}
      \centering
      \subfigure[]{\includegraphics[scale=0.48]{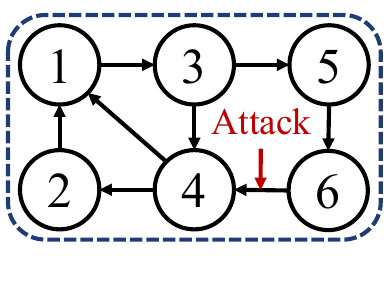}}
      \subfigure[]{\includegraphics[scale=0.48]{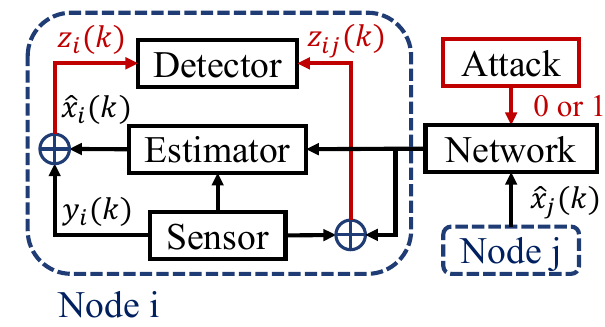}}
      \caption{System diagram: (a) the attacker tampers with partial channels of a distributed sensor network, (b) the internal architecture of each sensor node, which can construct two types of residues for attack detection.}
      \label{Fig_systemblock}
\end{minipage}
\end{figure}

It should be emphasized that for distributed estimation, each node can adopt its local estimate and neighbor information to generate two types of residues for attack detection. However, most previous works have not fully studied the security of distributed estimation under joint detection based on two types of residues. Besides, when the attacker only intrudes partial channels due to limited resources, the compromised local information of some nodes are directly exposed to their neighbors via the channels without being attacked (See Fig. \ref{Fig_systemblock}). In this case, the attack can still remain stealthy only if those compromised data can also bypass detection. However, to the best of our knowledge, few studies have investigated the security of distributed estimation under attacks that can keep stealthy by intruding partial channels. Motivated by these, we are interested in analyzing the security of distributed estimation in a more general attack scenario, where stealthy attacks can intrude partial channels to avoid detection of both types of residues. Moreover, we are committed to developing corresponding defense methods to ensure the system security. In contrast to existing literature such as \cite{zhou2022watermarking,zhou2022security}, the inherent vulnerability of distributed filtering considered in this work becomes more complex due to multiple constraints on attack stealthiness. Consequently, establishing the corresponding sufficient and necessary conditions poses greater mathematical challenges. The main contributions of this work are as follows:
\begin{itemize}
  \item [1)] Compared with \cite{liu2020false,yang2019distributed,zhou2022watermarking,zhou2022security,sui2021vulnerability,choraria2022design,ju2020distributed}, we study the vulnerability of distributed consensus filtering under a more stealthy and resource-saving attack, which can intrude partial channels to avoid the detection of both types of residues. Since the adversary needs to ensure that compromised local estimates can also bypass detection after being sent to their neighbors through normal channels, our results show that the security in this scenario further depends on the coupling of distributed estimation and the characteristics of system dynamics.
      \item [2)] In terms of security analysis, we first consider the worst-case that the attacker can intrude all channels to diverge the estimation error without being detected. The necessary and sufficient condition to achieve the above attack object is derived (\textbf{Theorem \ref{theorem 1}}), which is more stringent than the one in \cite{zhou2022watermarking,zhou2022security}. It is because that more residual information is utilized for attack detection. Then, we further analyze the insecurity of distributed estimation when only partial channels are attacked (\textbf{Theorem \ref{theorem 2}}).
        \item [3)] In terms of protection strategies, we first adopt the Euclidean distance between local estimates for detection. It is proved that this method can resist attacks in most cases, but still leaves a few security loopholes (\textbf{Theorem \ref{theorem 3}}). Thus, we further propose a coding-based defense method to enhance the detection capability (\textbf{Theorem \ref{theorem 4}}). 
            To balance the trade-off between security and coding costs, a procedure is also provided to select security-critical channels for encoding \textbf{(Algorithm \ref{algorithm 1})}.
\end{itemize}

The remainder of the paper is organized as follows. Section \ref{Section2} introduces the system framework. Section \ref{Section3} analyzes the insecurity of distributed estimator. Section \ref{Section4} provides two protection strategies and a procedure for selecting encoded channels. Finally, numerical simulation and some concluding remarks are given in Sections \ref{Section5} and \ref{Section6}, respectively.

\emph{Notations:} $\mathbb{R}$ is the set of real numbers, and $\mathbb{R}^n$ denotes the $n$-dimensional Euclidean space. For a matrix $\textsl{X}$, we define $\mathrm{rank}(\textsl{X})$, $\mathrm{tr}(\textsl{X})$, $\Vert \textsl{X}\Vert_2$, $\lambda_{\min}(\textsl{X})$, $\textsl{X}^T$, and $\textsl{X}^{-1}$ as its rank, trace, Euclidian norm, minimum eigenvalue, transpose, and inverse, respectively. $\mathrm{null}(\textsl{X})$ represents the null space of $\textsl{X}$. $\textsl{X} \geqslant 0$ (or $\textsl{X} > 0$) means that $\textsl{X}$ is positive semi-definite (or positive definite). The Kronecker product of matrixes $\textsl{X}$ and $\textsl{Y}$ is represented by $\textsl{X}\otimes \textsl{Y}$. $\mathrm{diag}(\textsl{X}_i)$ denotes the block-diagonal matrix with main diagonal elements $\textsl{X}_i$. $I_N$ is the $N$-dimensional identity matrix. $\theta_i$ denotes a vector of suitable dimension, whose $i$th element equals to $1$ and all the others are $0$. For a set $\textsl{S}$, its cardinality is defined as $\mathrm{card}(\textsl{S})$. $\mathbb{E}[\cdot]$ stands for the expectation of a random variable. $\mathcal{N}(\mu,\Sigma)$ refers to a Gaussian distribution with mean $\mu$ and covariance $\Sigma$.

\section{System Description}\label{Section2}
In this section, we will sequentially introduce the process, distributed estimator, detector, and attack model. The overall system architecture is shown in Fig. \ref{Fig_systemblock}.

\subsection{Process Model}\label{Section2.1}
We consider a discrete-time linear time-invariant (LTI) process whose mathematical model is described by
\begin{align}
x(k+1) = Ax(k)+w(k),\label{process}
\end{align}
where $A\in \mathbb{R}^{n\times n}$ is the state matrix, $x(k)\in \mathbb{R}^n$ denotes the process state, both the process noise $w(k)\in \mathbb{R}^n$ and the initial state $x(0)$ follow the zero-mean i.i.d. Gaussian distribution with covariances $Q \geq 0$ and $\Pi_0 \geq 0$, respectively. Besides, it is assumed that $w(k)$ is independent of $x(0)$. We employ a distributed sensor network consisting of $N$ sensors to jointly monitor $x(k)$. For the $i$th sensor node, its measurement equation is described as:
\begin{align}
y_i(k) = C_ix(k)+v_i(k), \label{measurement}
\end{align}
where $y_i(k)\in \mathbb{R}^{m_i}$ and $v_i(k)\in \mathbb{R}^{m_i}$ represent the sensor measurement and its noise, respectively. Assume that $v_i(k)$ is also zero-mean i.i.d. Gaussian with covariance $R_i>0$, and is independent of $x(0)$, $w(k)$, and $v_j(k), \forall i\neq j$ for all $k$. We adopt a directed graph $\mathcal{G}=(\mathcal{V},\mathcal{E})$ to describe the data transmission among the sensor network. Specifically, the set of nodes $\mathcal{V}=\{1,2,...,N\}$ and the set of edges $\mathcal{E} \subset \mathcal{V}\times \mathcal{V}$ denote the sensors and their communication channels, respectively. If the edge $(i,j) \in \mathcal{E}$, it indicates that there exists a communication channel connected from the $j$th sensor to the $i$th sensor. For the $i$th sensor, the set of its in-neighbors is defined as $\mathcal{N}_i=\{j:(i,j) \in \mathcal{E}\}$, whose cardinality is denoted by $d_i=\mathrm{card}(\mathcal{N}_i)$. Similarly, $\overline{\mathcal{N}}_i=\{j:(j,i) \in \mathcal{E}\}$ represents the set of its out-neighbors. The Laplacian matrix describing the topology with respect to the graph $\mathcal{G}$ is defined as $L\in \mathbb{R}^{N\times N}$.
\subsection{Distributed Estimator}\label{Section2.2}
To estimate the process state $x(k)$, the distributed sensor network adopts the distributed consensus estimator in \cite{yang2014stochastic,yang2017stochastic}. Define the local state estimate of the $i$th sensor as $\hat{x}_i(k)$, which is also the communication data broadcasted to its out-neighbors. Hence, after receiving $\hat{x}_j(k), \forall j \in \mathcal{N}_i$ from wireless channels, the $i$th sensor utilizes the following estimation algorithm to update its state estimate:
\begin{align}
\hat{x}_i(k+1) =& A\hat{x}_i(k)+K_i(k)[y_i(k)-C_i\hat{x}_i(k)] \nonumber\\
                & -\varepsilon A\sum\limits_{j\in \mathcal{N}_i}[\hat{x}_i(k)-\hat{x}_j(k)],\label{DKF}
\end{align}
where $K_i(k)\in \mathbb{R}^{n \times m_i}$ represents the estimator gain and the scalar $\varepsilon$ belongs to the range $(0, 1/\max_i(d_i))$ due to the requirement of consensus. Notice that the consensus term $\varepsilon$ can be further determined from the above feasible domain through the joint optimization approach in \cite{khan2022optimal}. We define the estimation error of the $i$th sensor as $e_i(k)=x(k)-\hat{x}_i(k)$ with covariance $P_{i}(k)=\mathbb{E}[e_i(k)e_i(k)^T]$. Besides, the cross covariance between the $i$th and $j$th sensors is denoted as $P_{ij}(k)=\mathbb{E}[e_i(k)e_j(k)^T]$. Then, based on \cite{yang2014stochastic,yang2017stochastic}, the optimal estimator gain to minimize the estimation error covariance $P_{i}(k)$ is $K_i^{*}(k) = A\{P_i(k)+\varepsilon\sum_{j\in N_i}[P_{ij}(k)-P_i(k)]\}C_i^T(C_iP_i(k)C_i^T+R_i)^{-1}$. Define the global estimation error of the entire sensor network as $e(k)=[e_1^T,...,e_N^T]^T$, whose covariance $P(k)=\mathbb{E}[e(k)e(k)^T]$ is a block matrix composed of $P_{i}(k)$ and $P_{ij}(k)$. It is proven that $P(k)$ can converge to the steady state under the following assumptions in \cite{yang2014stochastic,yang2017stochastic}:
\begin{assumption} \label{assumption1}
\rm The graph $\mathcal{G}$ is strongly connected.
\end{assumption}
\begin{assumption}\label{assumption2}
\rm $(A,Q^{1/2})$ is stabilizable.
\end{assumption}
\begin{assumption}\label{assumption3}
\rm $((I_N-\varepsilon L)\otimes A,{\rm diag}(C_i))$ is detectable, i.e., there exists a matrix $K$ in the form of $K=\mathrm{diag}(\bar{K}_i)$ such that $(I_N-\varepsilon)\otimes A-\mathrm{diag}(\bar{K}_i)\mathrm{diag}(C_i)$ is stable.
\end{assumption}

\begin{lemma} \label{lemma add 1}
Under Assumptions \ref{assumption1}-\ref{assumption3}, for any initial non-negative symmetric matrix $P(0)$, the estimation error covariance $P(k)$ of the distributed consensus filtering (3) is bounded for all $k$, and converges to a unique limit $\bar{P}>0$.
\end{lemma}

Hence, without loss of generality, we assume that (\ref{DKF}) has entered into the steady state at the initial time $k=0$. That is, $P(0)=\bar{P}>0$, whose $i$th diagonal block matrix is $\bar{P}_i$. Then, the steady-state estimator gain in (\ref{DKF}) can be rewritten as a fixed matrix as well, i.e., $K_i(k)=\bar{K}_i$.

\begin{remark}\label{remark_add}
\rm According to \cite{yang2014stochastic,yang2017stochastic}, Assumptions \ref{assumption1}, \ref{assumption2}, and \ref{assumption3} are the standard requirements to ensure the convergence and stability of distributed consensus filtering (\ref{DKF}). Notice that the detectability condition in Assumption \ref{assumption3} is weaker than the condition in \cite{olfati2009kalman}, which require $(A,C_i)$ to be locally detectable. However, it is stronger than the global detectability condition in existing works such as \cite{kamal2013information,battistelli2014consensus,wang2017convergence}, i.e., $(A,C)$ is detectable, where $C=[C_1^T,C_2^T,...,C_N^T]^T$. Notice that different from (\ref{DKF}), those in \cite{kamal2013information,battistelli2014consensus,wang2017convergence} need to broadcast information pairs (matrix-vector), both of which may be tampered with by attackers during transmission. Considering that  the joint detection of
information pairs is not yet mature, it implies that those estimation algorithms may be more vulnerable under cyber attacks. Besides, (\ref{DKF}) is one of the representative and basic distributed estimation algorithms, the methodology presented in subsequent sections for deriving necessary and sufficient conditions for its security vulnerabilities is highly instructive for other ones. It is one of our future works to study the commonalities of vulnerabilities among different distributed estimators.
\end{remark}

\subsection{False Data Detector}\label{Section2.3}
The $\chi^2$ detector is widely employed to diagnose data anomalies by examining the statistical properties of residues. In the distributed estimation, each sensor can utilize its local estimate $\hat{x}_i(k)$ and the received data  $\hat{x}_j(k), \forall j \in \mathcal{N}_i$ to construct two types of residues, i.e., $z_i(k)=y_i(k)-C_i\hat{x}_i(k)$ and $z_{ij}(k)=y_i(k)-C_i\hat{x}_j(k)$. According to \cite{zhou2022watermarking}, $z_i(k)$ and $z_{ij}(k)$ are zero-mean Gaussian vectors, whose steady-state covariances are $\Sigma_{i}=C_i\bar{P}_iC_i^T+R_i$ and $\Sigma_{ij}=C_i\bar{P}_jC_i^T+R_i$, respectively. We configure the $\chi^2$ detectors with the number of $1+d_i$ for the $i$th sensor, since the number of its in-neighbors is $d_i$. For the residue $z_{i}(k)$, the following hypothesis test is utilized as the detection criterion of the $\chi^2$ detector \cite{greenwood1996guide}:
\begin{align}
\sum_{s=k-J_i+1}^k [z_{i}(s)]^T\Sigma_{i}^{-1}z_{i}(s) \mathop{\stackrel{H_0}\lessgtr}_{H_1} \zeta_{i},\label{detector}
\end{align}
where $H_0$ is the null hypotheses indicating that $z_{i}(k)$ is normal, while $H_1$ is the opposite, $J_i$ and $\zeta_{i}$ are the window size and threshold, respectively. Similar to (\ref{detector}), $z_{ij}(k)$ can also be verified through the hypothesis test based on $\Sigma_{ij}$.

\subsection{Attack Model}\label{Section2.4}
Due to the vulnerability of wireless channels, malicious third parties may intercept and modify the transmitted data. It is assumed that the adversary acquires all system parameters including $A$, $Q$, $\varepsilon$, $L$, $C_i$, $R_i$, $\bar{K}_i$, $J_i$, and $\zeta_{i}, \forall i \in\mathcal{V}$, and is able to eavesdrop the transmitted data of each channel. Besides, we consider that the adversary can arbitrarily select several channels to launch attacks. Hence, we introduce a binary variable $\gamma_{ij}$, where $\gamma_{ij}=1$ means that the channel $(i,j)$ is under attack, while $\gamma_{ij}=0$ is the contrary. Define the local estimate of the $i$th sensor under attack as $\hat{x}_i^a(k), \forall i \in\mathcal{V}$. In what follows, the superscript ``a'' is utilized to denote the quantities under attack. Then, the model of the false data injection attack on the channel $(i,j)$ can be expressed as:
\begin{align}
\tilde{x}_{ij}^a(k)=\hat{x}_j^a(k)+\gamma_{ij}a_{ij}(k),\label{attack model}
\end{align}
where $\hat{x}_j^a(k)$ is the data transmitted by the $j$th sensor, $a_{ij}(k)\in \mathbb{R}^n$ is an arbitrary attack vector injected into the channel $(i,j)$, and $\tilde{x}_{ij}^a(k)$ is the data received by the $i$th sensor. In light of this, (\ref{attack model}) is a generalized attack framework that can encompass the mathematical representation of most integrity attacks in existing literature. On the basis of the Laplacian matrix $L$, we define an adjacency matrix $A_{\gamma}=[\gamma_{ij}]\in \mathbb{R}^{N\times N}$ to describe the attacked channels in the entire sensor network.

Recall that each sensor deploys several $\chi^2$ detectors for attack detection, and the residues under attack are $z_i^a(k)=y_i(k)-C_i\hat{x}_i^a(k)$, and $z_{ij}^a(k)=y_i(k)-C_i\tilde{x}_{ij}^a(k), \forall j \in \mathcal{N}_i, i \in\mathcal{V}$. Since the $\chi^2$ detector basically relies on the probability distributions of residues, the attack is strictly stealthy if and only if $z_i^a(k)$ and $z_{ij}^a(k)$ preserve the same statistical properties as $z_i(k)$ and $z_{ij}(k)$, respectively. Besides, similar to \cite{mo2010false,miao2016coding}, we consider the worst case that the adversary aims to corrupt the estimator (\ref{DKF}) by diverging its estimation error to infinity. In the following, we define the security of the distributed estimator (\ref{DKF}) and summarize the objects of the attacker (\ref{attack model}).

\begin{definition}\label{definitionl}
\rm If there exists at least one matrix $A_{\gamma}$ and a corresponding sequence $a_{ij}(k),\forall j \in \mathcal{N}_i, i \in\mathcal{V}$ such that
 \begin{itemize}
 \item [1)] All residues of sensor networks keep strictly stealthy, i.e., $z_i^a(k)\sim\mathcal{N}(0,\Sigma_{i})$, $z_{ij}^a(k)\sim\mathcal{N}(0,\Sigma_{ij}), \forall j \in \mathcal{N}_i, i \in\mathcal{V}$.
 \item [2)] The estimation error under attack diverges to infinity over time, i.e., $\lim_{k\to\infty}\Vert e^a(k)\Vert_2\to\infty$,
 \end{itemize}
 then the distributed consensus filter (\ref{DKF}) is called insecure \cite{zhou2022watermarking} (or perfectly attackable \cite{mo2010false}).
\end{definition}

\begin{remark}\label{remarkl}
\rm In \cite{zhou2022watermarking}, the authors proposed a similar attack model $\tilde{x}_{ij}^a(k)=\hat{x}_j^a(k)+a_{ij}(k)$, which requires all channels to be attacked. On the one hand, due to the limitation of attack resource, the adversary cannot tamper with all channels especially in large-scale sensor networks. On the other hand, when $\gamma_{ij}=1$, (\ref{attack model}) can be simplified into $\tilde{x}_{ij}^a(k)=\hat{x}_j^a(k)+a_{ij}(k)$. Hence, (\ref{attack model}) is an extended version of the one in \cite{zhou2022watermarking}, and can describe which channels are attacked in a more intuitive and clear way. Moreover, it is worth noting that when $\gamma_{ij}=0$, the true value of $\hat{x}_j^a(k)$ are directly exposed to the $i$th sensor. It indicates that different from \cite{zhou2022watermarking}, the adversary with the attack model (\ref{attack model}) needs to further design its strategy from a global perspective of the sensor network to avoid $\hat{x}_j^a(k), \gamma_{ij}=0$ being detected. Besides, since $z_{i}^a(k)$ is also the information available to each sensor for attack detection, we consider a more general scenario that the attack is designed to bypass the detection of both $z_{ij}^a(k)$ and $z_{i}^a(k)$.
\end{remark}

\subsection{Problems of Interest}\label{Section2.5}
In terms of security analysis and defense strategies, we are mainly interested in the following three problems:
 \begin{itemize}
 \item [1)] For the attack (\ref{attack model}) that tampers with partial channels, what is the necessary and sufficient condition to achieve its attack objects shown in Definition \ref{definitionl}?
 \item [2)] When the coding scheme is taken as a defense strategy, how to design coding matrixes for different channels?
 \item [3)] Can we deploy the coding scheme on partial channels to balance the trade-off between security and cost? If so, which channels should be prioritized for encoding?
 \end{itemize}

 We will present the main results for the above problems in the following two sections.
\section{Security Analysis}\label{Section3}
This section investigates the circumstances under which the distributed estimator (\ref{DKF}) is insecure in the presence of the attack (\ref{attack model}). Specifically, we first consider the same case of \cite{zhou2022watermarking}, i.e., all channels are attacked. By comparison, we reveal the additional constraints imposed by the residual $z_{i}^a(k)$ on the attack. Then, we extend this case to a more general scenario where partial channels are under attack.

\subsection{Scenario I: all channels are under attack}\label{Section3.1}
 For the $i$th sensor, the iterative equation (\ref{DKF}) of its local state estimate under the attack (\ref{attack model}) can be rewritten as:
\begin{align}
\hat{x}_i^a(k+1) =& A\hat{x}_i^a(k)+\bar{K}_i[y_i(k)-C_i\hat{x}_i^a(k)] \nonumber\\
                & -\varepsilon A\sum\limits_{j\in \mathcal{N}_i}[\hat{x}_i^a(k)-\tilde{x}_{ij}^a(k)],\label{DKF_attack}
\end{align}
where $\tilde{x}_{ij}^a(k)=\hat{x}_j^a(k)+\gamma_{ij}a_{ij}(k)$ is the data received from the channel $(i,j)$. Define the difference between the normal system and the compromised one as $\Delta\hat{x}_i^a(k)=\hat{x}_i^a(k)-\hat{x}_i(k)$. By subtracting (\ref{DKF}) from (\ref{DKF_attack}), we have
\begin{align}
\Delta\hat{x}_i^a(k+1) =& [(1-\varepsilon d_i)A-\bar{K}_iC_i]\Delta\hat{x}_i^a(k) \nonumber\\
                        & +\varepsilon A\sum\limits_{j\in \mathcal{N}_i}[\Delta\hat{x}_j^a(k)+\gamma_{ij}a_{ij}(k)],\label{DKF_attack_difference}
\end{align}
where $\Delta\hat{x}_i^a(0)=0$. Similarly, the differences on residues are defined as $\Delta z_i^a(k)=z_i(k)-z_i^a(k)$ and $\Delta z_{ij}^a(k)=z_{ij}(k)-z_{ij}^a(k)$, which equal to $C_i\Delta\hat{x}_i^a(k)$ and $C_i[\Delta\hat{x}_j^a(k)+\gamma_{ij}a_{ij}(k)]$, respectively. As mentioned in Remark \ref{remarkl}, when all channels are attacked, i.e., $\gamma_{ij}=1, \forall j \in \mathcal{N}_i, i \in\mathcal{V}$, each node cannot receive the true value of $\hat{x}_j^a(k)$ due to the isolation of attack signals. In other words, each node in the sensor network can be regarded as an independent information silo, because its state estimation and attack detection are independent of each other. Besides, the second condition of Definition \ref{definitionl} indicates that there exists at least one sensor with infinite estimation error, i.e., $\exists i \in\mathcal{V}, \lim_{k\to\infty}\Vert e_i^a(k)\Vert_2\to\infty$. It is equivalent to $\lim_{k\to\infty}\Vert \Delta\hat{x}_i^a(k)\Vert_2\to\infty$, since $\mathbb{E}[\Vert \Delta\hat{x}_i^a(k)-e_i^a(k)\Vert_2]=\sqrt{\mathrm{tr}(\bar{P}_i)}$. In summary, the attack objects in this case can be transformed into: 1) $\exists i \in\mathcal{V}, \forall j \in \mathcal{N}_i, z_i^a(k)\sim\mathcal{N}(0,\Sigma_{i}), z_{ij}^a(k)\sim\mathcal{N}(0,\Sigma_{ij})$, and 2) $\lim_{k\to\infty}\Vert \Delta\hat{x}_i^a(k)\Vert_2\to\infty$. In the following, we derive the necessary and sufficient condition for the attack (\ref{attack model}) to achieve the above objects.
\begin{theorem} \label{theorem 1}
Under Assumptions \ref{assumption1}-\ref{assumption3}, for the $i$th sensor, the attack (\ref{attack model}) can diverge the estimation error of the distributed estimator (\ref{DKF}) without triggering the alarm of the detector (\ref{detector}), if and only if 1) $\mathrm{rank}(C_i)<n$, and 2) there exists at least one nonzero vector $x\in \mathbb{R}^{l_i}$ such that $\mathrm{rank}(A\Xi^i)=\mathrm{rank}([A\Xi^i,\Xi^ix])$, where $\Xi^i=[\sigma_1^i,...,\sigma_{l_i}^i]$, $l_i=n-\mathrm{rank}(C_i)$, and $\sigma_s^i\in \mathrm{null}(C_i), s=1,...,l_i$ are linearly independent of each other.

\noindent\textbf{Proof.} \rm We first prove the necessity. There are four cases for the non-homogeneous linear equation $C_i\Delta\hat{x}_i^a(k)=\Delta z_i^a(k)$ with respect to $\Delta\hat{x}_i^a(k)$. Specifically, if $\mathrm{rank}(C_i)=n=m_i$, $C_i$ is an invertible matrix such that $\Delta\hat{x}_i^a(k)=(C_i)^{-1}\Delta z_i^a(k)\triangleq \rho_1^i(k)$. When $\mathrm{rank}(C_i)=n<m_i$, we can utilize an elementary transformation matrix $D_i\in \mathbb{R}^{m_i\times m_i}$ such that $D_iC_i=[\tilde{C}_i^T,0^T]^T$, where $\mathrm{rank}(\tilde{C}_i)=n$. Then, we can obtain that $\tilde{C}_i\Delta\hat{x}_i^a(k)=\tilde{D}_i\Delta z_i^a(k)$, where $\tilde{D}_i$ is a sub-matrix of $D_i$ from the $1$st row to the $n$th row. It yields that $\Delta\hat{x}_i^a(k)=(\tilde{C}_i)^{-1}\tilde{D}_i\Delta z_i^a(k)\triangleq \rho_2^i(k)$. In the case of $\mathrm{rank}(C_i)=m_i<n$, the nontrivial solutions of $C_i\Delta\hat{x}_i^a(k)=0$ are denoted by $\mathrm{null}(C_i)=\mathrm{span}\{\sigma_1^i,...,\sigma_{l_i}^i\}$. Hence, $\Delta\hat{x}_i^a(k)$ can be rewritten as $\Delta\hat{x}_i^a(k)=\sum_{s=1}^{l_i}\alpha_s^i(k)\sigma_s^i+\rho_3^i(k)$, where $\rho_3^i(k)$ is the minimum norm solution of $C_i\Delta\hat{x}_i^a(k)=\Delta z_i^a(k)$. Note that $C_i$ is row full rank in this case, and thus $C_iC_i^T$ is positive defined and invertible. According to the least squares, we have $\rho_3^i(k)=C_i^T(C_iC_i^T)^{-1}\Delta z_i^a(k)$. Finally, for $\mathrm{rank}(C_i)<n$ and $\mathrm{rank}(C_i)<m_i$, we can also derive that $\Delta\hat{x}_i^a(k)=\sum_{s=1}^{l_i}\alpha_s^i(k)\sigma_s^i+\rho_4^i(k)$, where $\rho_4^i(k)=\tilde{C}_i^T(\tilde{C}_i\tilde{C}_i^T)^{-1}\tilde{D}_i\Delta z_i^a(k)$, and $\mathrm{rank}(\tilde{C}_i)=\mathrm{rank}(C_i)$.

The attack needs to maintain $z_i^a(k)\sim\mathcal{N}(0,\Sigma_{i})$ to keep strictly stealthy. Hence, in the condition that $\mathrm{rank}(C_i)=n=m_i$, we can apply the triangle inequality to obtain that
\begin{align}
\mathbb{E}[\Vert \rho_1^i(k)\Vert_2]&\leq \mathbb{E}[\Vert(C_i)^{-1}\Vert_2(\Vert z_i(k)\Vert_2+\Vert z_i^a(k)\Vert_2)]\nonumber\\
&=2\Vert (C_i)^{-1}\Vert_2 \sqrt{\mathrm{tr}(\Sigma_{i})}\triangleq \beta_1^i. \label{appendix1_equation1}
\end{align}
Similarly, for the remaining three cases, we can deduce that $\mathbb{E}[\Vert \rho_2^i(k)\Vert_2]\leq 2\Vert (\tilde{C}_i)^{-1}\tilde{D}_i\Vert_2\sqrt{\mathrm{tr}(\Sigma_{i})}\triangleq \beta_2^i$, $\mathbb{E}[\Vert \rho_3^i(k)\Vert_2]\leq 2\Vert C_i^T(C_iC_i^T)^{-1}\Vert_2\sqrt{\mathrm{tr}(\Sigma_{i})}\triangleq \beta_3^i$, and $\mathbb{E}[\Vert \rho_4^i(k)\Vert_2]\leq 2\Vert \tilde{C}_i^T(\tilde{C}_i\tilde{C}_i^T)^{-1}\tilde{D}_i \Vert_2\sqrt{\mathrm{tr}(\Sigma_{i})}\triangleq \beta_4^i$. Clearly, if $\mathrm{rank}(C_i)=n$, the expectation of $\Vert\Delta\hat{x}_i^a(k)\Vert_2$ is bounded by $\beta_1^i$ or $\beta_1^2$, which contradicts $\lim_{k\to\infty}\Vert \Delta\hat{x}_i^a(k)\Vert_2\to\infty$. On the contrary, when $\mathrm{rank}(C_i)=m_i<n$, one has
\begin{align}
\mathbb{E}[\Vert\Delta\hat{x}_i^a(k)\Vert_2]&\geq \Vert\sum_{s=1}^{l_i}\alpha_s^i(k)\sigma_s^i\Vert_2-\mathbb{E}[\Vert\rho_3^i(k)\Vert_2]\nonumber\\
                                   &\geq \Vert\Xi^i \alpha^i(k)\Vert_2-\beta_3^i\nonumber\\
                                   &\geq \sqrt{\lambda_{min}([\Xi^i]^T\Xi^i)}\Vert\alpha^i(k)\Vert_2-\beta_3^i ,\label{appendix1_equation_add}
\end{align}
where $\alpha^i(k)=[\alpha_1^i(k),...,\alpha_{l_i}^{i}(k)]^T$, and $\lambda_{min}([\Xi^i]^T\Xi^i)>0$ since $\Xi^i$ is full column rank such that $[\Xi^i]^T\Xi^i$ is positive define. Thus, the attack can diverge $\Delta\hat{x}_i^a(k)$ by choosing $\Vert\alpha^i(k)\Vert_2\to \infty$. Note that the case of $\mathrm{rank}(C_i)<n$ and $\mathrm{rank}(C_i)<m_i$ is also the same. In a word, the attack can diverge $\Delta\hat{x}_i^a(k)$ only if $\mathrm{rank}(C_i)<n$. Then, we have $\Delta\hat{x}_i^a(k)=\Xi^i \alpha^i(k)+\rho^i(k)$, where $\mathbb{E}[\Vert \rho^i(k)\Vert_2]\leq \beta^i$, and $\{\rho^i(k),\beta^i\}=\{\rho_3^i(k),\beta_3^i\}$ or $\{\rho_4^i(k),\beta_4^i\}$. Similarly, when $\gamma_{ij}=1$, one has $\Delta\hat{x}_j^a(k)+a_{ij}(k)=\Xi^i \alpha^{ij}(k)+\rho^{ij}(k)$, where $\alpha^{ij}(k)$ and $\rho^{ij}(k)$ have the same meaning as $\alpha^{i}(k)$ and $\rho^{i}(k)$, respectively. Moreover, $\mathbb{E}[\Vert \rho^{ij}(k)\Vert_2]$ is also bounded.

Besides, $\Delta\hat{x}_i^a(k+1)$ should also follow $\Delta\hat{x}_i^a(k+1)=\Xi^i \alpha^i(k+1)+\rho^i(k+1)$. Hence, based on (\ref{DKF_attack_difference}), one has
\begin{align}
\Xi^i \alpha^i(k+&1) = [(1-\varepsilon d_i)A-\bar{K}_iC_i][\Xi^i\alpha^i(k)+\rho^i(k)] \nonumber\\
                        &+\varepsilon A\sum\limits_{j\in \mathcal{N}_i}[\Xi^i \alpha^{ij}(k)+\rho^{ij}(k)]-\rho^i(k+1)\nonumber\\
                    =& A\Xi^i[(1-\varepsilon d_i)\alpha^i(k)+\varepsilon\sum\limits_{j\in \mathcal{N}_i}\alpha^{ij}(k)]+[(1-\varepsilon d_i)\nonumber\\
                    &A-\bar{K}_iC_i]\rho^i(k)+\varepsilon A\sum\limits_{j\in \mathcal{N}_i}\rho^{ij}(k)-\rho^i(k+1)\nonumber\\
                    \triangleq& A\Xi^i\tilde{\alpha}^i(k)+\tilde{\rho}^i(k),\label{appendix1_equation2}
\end{align}
where the second equality is based on $C_i\Xi^i=0$. Note that
\begin{align*}
\mathbb{E}[\Vert \tilde{\rho}^i(k)\Vert_2]\leq& \Vert(1-\varepsilon d_i)A-\bar{K}_iC_i\Vert_2 \mathbb{E}[\Vert \rho^i(k)\Vert_2]+\varepsilon\Vert A\Vert_2\nonumber\\
                                  &\sum\limits_{j\in \mathcal{N}_i}\mathbb{E}[\Vert\rho^{ij}(k)\Vert_2]+\mathbb{E}[\Vert \rho^i(k+1)\Vert_2]\nonumber\\
                                   \leq & [1+\Vert(1-\varepsilon d_i)A-\bar{K}_iC_i\Vert_2]\beta^i+\varepsilon\Vert A\Vert_2\beta^{ij}.
\end{align*}
Besides, the attack object $\lim_{k\to\infty}\Vert \Delta\hat{x}_i^a(k)\Vert_2\to\infty$ is equivalent to $\lim_{k\to\infty}\Vert\alpha^i(k+1)\Vert_2\to \infty$. Thus, one can deduce that $\lim_{k\to\infty}[\Xi^i\frac{\alpha^i(k+1)}{\Vert\alpha^i(k+1)\Vert_2}-A\Xi^i\frac{\tilde{\alpha}^i(k)}{\Vert\alpha^i(k+1)\Vert_2}]=\lim_{k\to\infty}[\frac{\tilde{\rho}^i(k)}{\Vert\alpha^i(k+1)\Vert_2}]\to 0$. In other words, $A$ and $\Xi^i$ should satisfy: $\exists x,y\in \mathbb{R}^n\neq 0$, $\Xi^ix-A\Xi^iy=0$. It also means that $\exists x\in \mathbb{R}^n\neq 0$, $\mathrm{rank}(A\Xi^i)=\mathrm{rank}([A\Xi^i,\Xi^ix])$.

Next, we prove the sufficiency. The condition $\mathrm{rank}(C_i)<n$ indicates that $\Xi^i\neq 0$, while the second one means that there exists $\{x^*,y^*\}$ such that $\Xi^ix^*=A\Xi^iy^*$. By induction, we design the attack at time $k=0$ as $a_{ij}(0)=-\Delta\hat{x}_j^a(0)+\eta(0)\Xi^iy^*, \forall j\in \mathcal{N}_i$, where $\eta(0)\in \mathbb{R}$ is arbitrarily chosen. From the definition of $\Delta z_{ij}^a(k)$, one has $z_{ij}^a(0)=z_{ij}(0)-\eta(0)C_i\Xi^iy^*=z_{ij}(0)$. Since $\Delta\hat{x}_i^a(0)=0$, we have $z_{i}^a(0)=z_{i}(0)-C_i\Delta\hat{x}_i^a(0)=z_{i}(0)$. Thus, the attack is strictly stealthy at time $k=0$. Based on (\ref{DKF_attack_difference}), $\Delta\hat{x}_i^a(1)$ can be expressed as
\begin{align}
\Delta\hat{x}_i^a(1)=&[(1-\varepsilon d_i)A-\bar{K}_iC_i]\Delta\hat{x}_i^a(0)+\varepsilon A\sum\limits_{j\in \mathcal{N}_i}\eta(0)\Xi^iy^*\nonumber\\
=&\varepsilon d_i\eta(0)A\Xi^iy^*=\varepsilon d_i\eta(0)\Xi^ix^*. \label{appendix1_equation3}
\end{align}
At time $k=1$, the attack is designed as $a_{ij}(1)=-\Delta\hat{x}_j^a(1)+\Xi^i[\eta(1)y^*-(1-\varepsilon d_i)\eta(0)x^*], \forall j\in \mathcal{N}_i$. Then, we have $z_{ij}^a(1)=z_{ij}(1)-C_i\Xi^i[\eta(1)y^*-(1-\varepsilon d_i)\eta(0)x^*]=z_{ij}(1)$ and  $z_{i}^a(1)=z_{i}(1)-\varepsilon d_i\eta(0)C_i\Xi^ix^*=z_{i}(1)$. Hence, the attack also keeps strictly stealthy. Accordingly, $\Delta\hat{x}_i^a(2)$ can be written as
\begin{align}
\Delta\hat{x}_i^a(2)=&[(1-\varepsilon d_i)A-\bar{K}_iC_i][\varepsilon d_i\eta(0)\Xi^ix^*]+\varepsilon A\sum\limits_{j\in \mathcal{N}_i}\Xi^i\nonumber\\
                     &[\eta(1)y^*-(1-\varepsilon d_i)\eta(0)x^*]\nonumber\\
                    =&(1-\varepsilon d_i)\varepsilon d_i\eta(0)A\Xi^ix^*+\varepsilon d_i\eta(1)A\Xi^iy^*-\nonumber\\
                     & (1-\varepsilon d_i)\varepsilon d_i\eta(0)A\Xi^ix^*=\varepsilon d_i\eta(1)\Xi^ix^*. \label{appendix1_equation4}
\end{align}
At time $k=2,3,...$, the attack can also construct the similar strategy. By choosing $\eta(k)\to\infty$, $\Delta\hat{x}_i^a(k)$ can be diverged at any time $k$. The proof is thus completed.
 \hfill\rule{2mm}{2mm}
\end{theorem}

The condition of $\mathrm{rank}(C_i)<n$ is required for the attacker, since the component of $\Delta\hat{x}_i^a(k)$ that tends to infinity depends on the null space of $C_i$, i.e., $\mathrm{null}(C_i)$. Notice that for the measurement matrix $C_i$, the number of its rows is generally less than the number of its columns, i.e., $m_i<n$. Hence, the attack is hardly restricted by this condition. Besides, when only $z_{ij}^a(k)$ is adopted to detect attacks, it shows in \cite{zhou2022watermarking} that the attacker only requires to satisfy $\mathrm{rank}(C_i)<n$. Once both $z_{ij}^a(k)$ and $z_{i}^a(k)$ are exploited by the detector, the adversary needs to further ensure that $A$ and $C_i$ satisfy an extra condition in Theorem \ref{theorem 1}. It is because that the attack can bypass the detection of $z_{i}^a(k)$, only if the infinite component of $\Delta\hat{x}_i^a(k)$ still belongs to $\mathrm{null}(C_i)$ before and after the iterative recursion in (\ref{DKF_attack_difference}). That is, there exists $\Xi^iy, y\neq 0$ that can be converted to $\Xi^ix, x\neq 0$ through the linear mapping of $A$. Specifically, when $\mathrm{rank}(C_i)=n-1$, the above condition is equivalent to $\exists x\in \mathbb{R}\neq 0$, $A\Xi^i=x\Xi^i$, where $\Xi^i\in \mathbb{R}^{n}$. It indicates that the estimator is insecure only if $\Xi^i$ is also an eigenvector of $A$ in this case. Besides, when $A=I_n$, the above condition always holds, and thus does not restrict the attacker. Furthermore, in proving the sufficiency of Theorem \ref{theorem 1}, we provide a feasible attack strategy for generating the false data $a_{ij}(k)$ injected into each channel $(i,j)$. It illustrates that the attack objects in Definition \ref{definitionl} are achievable by the adversary.

Note that the second condition of Theorem \ref{theorem 1} is equivalent to $\exists x\neq 0$, $\Xi^ix-A\Xi^iy=0$ in terms of $y$ is solvable. Hence, its sufficient condition is that the matrix $[\Xi^i,A\Xi^i]$ with $2l_i$ columns is not full column rank. On the contrary, if the column rank of $[\Xi^i,A\Xi^i]$ equals to $2l_i$, the second condition of Theorem \ref{theorem 1} must not be satisfied. Moreover, if any one of the necessary and sufficient conditions in Theorem \ref{theorem 1} is not satisfied, the strictly stealthy attack (\ref{attack model}) can only yield bounded estimation error. In the following lemma, an upper bound of $\Delta\hat{x}_i^a(k)$ is derived to quantify the estimation performance degradation in this case.

\begin{lemma} \label{lemma 1}
Under Assumptions \ref{assumption1}-\ref{assumption3}, when any condition in Theorem \ref{theorem 1} is not satisfied, the estimation error of the $i$th sensor under the strictly stealthy attack (\ref{attack model}) is bounded by $\mathbb{E}[\Vert\Delta\hat{x}_i^a(k)\Vert_2]\leq 2\Vert \tilde{C}_i^T(\tilde{C}_i\tilde{C}_i^T)^{-1}\tilde{D}_i \Vert_2\sqrt{\mathrm{tr}(\Sigma_{i})}$, where $\tilde{C}_i=C_i$ and $\tilde{D}_i=I_{m_i}$ if $\mathrm{rank}(C_i)=m_i$. Otherwise, if $\mathrm{rank}(C_i)<m_i$, $\tilde{C}_i\in \mathbb{R}^{\mathrm{rank}(C_i)\times n}$ and $\tilde{D}_i\in \mathbb{R}^{\mathrm{rank}(C_i)\times m_i}$ are two constant matrixes such that $\mathrm{rank}(\tilde{C}_i)=\mathrm{rank}(C_i)$ and $\tilde{D}_iC_i=\tilde{C}_i$.

\noindent\textbf{Proof.} The proof is shown in Appendix \ref{appendix2}.
 \hfill\rule{2mm}{2mm}
\end{lemma}

According to Lemma \ref{lemma 1}, a countermeasure is to configure sensors such that $\forall i \in\mathcal{V}$, $C_i$ does not satisfy the second condition in Theorem \ref{theorem 1}. Then, the attacker can only cause limited estimation performance degradation.

\subsection{Scenario II: partial channels are under attack}\label{Section3.2}
When some channels $(s,i), s \in \overline{\mathcal{N}}_i$ are not attacked, i.e., $\gamma_{si}=0$, the corresponding out-neighbors of the $i$th sensor can receive the true value of $\Delta\hat{x}_i^a(k)$. Hence, the attack needs to guarantee that $\Delta\hat{x}_i^a(k)$ can bypass the detection of those out-neighbors as well. To be specific, when $\gamma_{si}=0$, the $s$th sensor can generate the residue $z_{si}^a(k)=y_s(k)-C_s\hat{x}_i^a(k)$. That is, $C_s\Delta\hat{x}_i^a(k)=z_{si}(k)-z_{si}^a(k)$, where $z_{si}^a(k)\sim\mathcal{N}(0,\Sigma_{si})$ to keep stealthy. The channels without being attacked can be described as $A_{L}-A_{\gamma}$, where $A_{L}$ is the adjacency matrix of $L$. Then, based on the $i$th column of $A_{L}-A_{\gamma}$, we can stack $C_s, \gamma_{si}=0, \forall s \in \overline{\mathcal{N}}_i$ and $C_i$ into a column, i.e., $\tilde{C}_i^a\triangleq [...,(C_s)^T...,(C_i)^T]^T$. It implies that the infinite component of $\Delta\hat{x}_i^a(k)$ is determined by the null space of $\tilde{C}_i^a$, i.e., $\mathrm{null}(\tilde{C}_i^a)=\mathrm{span}\{\tilde{\sigma}_1^i,...,\tilde{\sigma}_{\tilde{l}_i}^i\}$, where $\tilde{l}_i=n-\mathrm{rank}(\tilde{C}_i^a)$, and $\tilde{\sigma}_t^i\in \mathrm{null}(\tilde{C}_i^a), t=1,...,\tilde{l}_i$.

Clearly, $\mathrm{null}(\tilde{C}_i^a)$ is the intersection of $\mathrm{null}(C_s), \gamma_{si}=0, \forall s \in \overline{\mathcal{N}}_i$ and $\mathrm{null}(C_i)$, and thus $\mathrm{null}(\tilde{C}_i^a)\subset \mathrm{null}(C_i)$. Besides, $\mathrm{rank}(\tilde{C}_i^a)$ increases monotonically as the number of $C_s$ increases. When $\mathrm{rank}(\tilde{C}_i^a)=n$, the attacker cannot diverge the estimation error. We stack $\Delta\hat{x}_i^a(k), \forall i \in\mathcal{V}$ into a column, i.e., $\Delta\hat{x}^a(k)=[(\Delta\hat{x}_1^a(k))^T,...,(\Delta\hat{x}_N^a(k))^T]^T$. Similarly, $\Vert e^a(k)\Vert_2\to\infty$ is equivalent to $\Vert \Delta\hat{x}^a(k)\Vert_2\to\infty$. For the adversary that attacks partial channels, the following theorem derives the necessary and sufficient condition to realize its attack objects in Definition \ref{definitionl}.
\begin{theorem} \label{theorem 2}
Under Assumptions \ref{assumption1}-\ref{assumption3}, when partial channels are subject to the attack (\ref{attack model}), the adversary can keep strictly stealthy and destabilize $\Delta\hat{x}^a(k)$, if and only if 1) $\exists i \in\mathcal{V}, \mathrm{rank}(\tilde{C}_i^a)<n$, and 2) there exists at least one sequence $\{\check{\alpha}(k)\}$ such that the following equation has the nontrivial solution $\alpha(k+1)$:
\begin{align}
\mathrm{diag}(\tilde{\Xi}^i) \alpha(k+1)=\Phi_1^\gamma\alpha(k)+\Phi_2^\gamma\check{\alpha}(k),\label{theorem2_equation1}
\end{align}
where $\tilde{\Xi}^i=[\tilde{\sigma}_1^i,...,\tilde{\sigma}_{\tilde{l}_i}^i]$, $\Phi_1^\gamma\triangleq [I_N-\varepsilon(L+A_{\gamma})]\otimes A \mathrm{diag}(\tilde{\Xi}^i)$, $\Phi_2^\gamma\triangleq \varepsilon \mathrm{diag}(A_{\gamma}^{[i]}\otimes (A\Xi^i))$, and $A_{\gamma}^{[i]}$ is the $i$th row of $A_{\gamma}$.

\noindent\textbf{Proof.} The proof is shown in Appendix \ref{appendix3}.
 \hfill\rule{2mm}{2mm}
\end{theorem}

The $i$th column (row) of $A_{\gamma}$ indicates the attacked channels between the $i$th sensor and its out-neighbors (in-neighbors). As mentioned before, the infinite component of $\Delta\hat{x}_i^a(k)$ depends on the $i$th column of $A_{\gamma}$. From (\ref{theorem2_equation1}), the $i$th row of $A_{\gamma}$ determines whether this infinite component does not trigger the alarm at time $k+1$ after the iteration of (\ref{DKF_attack_difference}). Moreover, if any condition in Theorem \ref{theorem 2} is not satisfied, the global estimation error $\Delta\hat{x}^a(k)$ under the strictly stealthy attack (\ref{attack model}) must be bounded. Similar to Lemma \ref{lemma 1}, one can derive an upper bound of $\mathbb{E}[\Vert\Delta\hat{x}^a(k)\Vert_2]$ as well.

For a given $\alpha(k)$, if there does not exist $\check{\alpha}(k)$ to implement the iteration in (\ref{theorem2_equation1}), the attacker will either trigger the alarm at time $k+1$ or be unable to maintain $\Vert \Delta\hat{x}^a(k+1)\Vert_2\to\infty$. In other words, we need to search the sequence $\{\check{\alpha}(k)\}$ over the entire time domain to determine the security of distributed estimator (\ref{DKF}), which may consume a lot of time and computing resources. In view of this, we further derive a simple form of (\ref{theorem2_equation1}) for the special case in the following lemma.
\begin{lemma} \label{lemma 2}
Under Assumptions \ref{assumption1}-\ref{assumption3}, for $\forall s\in\overline{\mathcal{N}}_i, \gamma_{si}=0$, when at least one channel $(s,t)$ is attacked, i.e., $\exists t \in \mathcal{N}_s$, $\gamma_{st}=1$, the attack (\ref{attack model}) can diverge $\Delta\hat{x}^a_i(k)$ of distributed estimator (\ref{DKF}) and bypass all detectors (\ref{detector}), if and only if, 1) $\mathrm{rank}(\tilde{C}_i^a)<n$, and 2) $\exists x\in \mathbb{R}^{\tilde{l}_i}\neq 0$ such that $\mathrm{rank}(A\Xi^i)=\mathrm{rank}([A\Xi^i,\tilde{\Xi}^ix])$.

\noindent\textbf{Proof.} \rm We begin with the necessity. When $\gamma_{ij}=0$, the $i$th sensor can obtain $\Delta\hat{x}_j^a(k)$, which satisfies $C_i\Delta\hat{x}_j^a(k)=z_{ij}(k)-z_{ij}^a(k)$. Hence, we have $\tilde{C}_j^a\triangleq [...,(C_i)^T...,(C_j)^T]^T$, which means that $\mathrm{null}(\tilde{C}_j^a)\subset \mathrm{null}(C_i)$ if $\gamma_{ij}=0$. Thus, there must exist $\check{\tilde{\alpha}}^j(k)\in \mathbb{R}^{l_i}$ such that $\Xi^i\check{\tilde{\alpha}}^j(k)=\tilde{\Xi}^j \alpha^j(k)$ in (\ref{appendix3_equation1}). Similarly, since $\mathrm{null}(\tilde{C}_i^a)\subset \mathrm{null}(C_i)$, $\tilde{\Xi}^i \alpha^i(k)$ can be rewritten as $\Xi^i\check{\tilde{\alpha}}^i(k)$. Then, we can simplify (\ref{appendix3_equation1}) into
\begin{align}
\tilde{\Xi}^i \alpha^i&(k+1)=A\Xi^i\{(1-\varepsilon d_i)\check{\tilde{\alpha}}^i(k)+\varepsilon\sum\limits_{j\in \mathcal{N}_i}[(1-\gamma_{ij})\check{\tilde{\alpha}}^j(k)\nonumber\\
                            &+\gamma_{ij}\alpha^{ij}(k)]\}+\tilde{\rho}^i(k)  \triangleq A\Xi^i\tilde{\alpha}^i(k)+\tilde{\rho}^i(k),  \label{appendix4_equation1}
\end{align}
where if $\exists \gamma_{ij}=1$, $\tilde{\alpha}^i(k)\in \mathbb{R}^{l_i}$ can be freely and arbitrarily determined by the attacker via $\alpha^{ij}(k)$, and thus is the same as the one in (\ref{appendix1_equation2}). When partial channels are attacked, $\Delta\hat{x}_i^a(k)$ can be received by some out-neighbors of the $i$th sensor, i.e., $\forall s\in\overline{\mathcal{N}}_i, \gamma_{si}=0$. However, if $\exists t \in \mathcal{N}_s$, $\gamma_{st}=1$, it means that (\ref{appendix4_equation1}) also holds for the $s$th sensor, such that the evolution of $\alpha^s(k)$ is governed by $\alpha^{st}(k)$, rather than $\alpha^i(k)$. In other words, (\ref{appendix3_equation2}) can be decoupled into (\ref{appendix4_equation1}) for the $i$th sensor. Then, similar to Theorem \ref{theorem 2}, $\mathrm{rank}(\tilde{C}_i^a)<n$ is required to ensure $\Vert\Delta\hat{x}_i^a(k)\Vert_2\to\infty$. Besides, similar to (\ref{appendix1_equation2}), $A$, $\Xi^i$, and $\tilde{\Xi}^i$ should satisfy $\mathrm{rank}(A\Xi^i)=\mathrm{rank}([A\Xi^i,\tilde{\Xi}^ix])$.

The sufficiency is proved by induction. The conditions of Lemma \ref{lemma 2} mean that $\exists x^*,y^*\neq0$, $\tilde{\Xi}^ix^*=A\Xi^iy^*$. At time $k=0$, the attacker tampers with the channel $(i,j)$ based on $a_{ij}(0)=\eta(0)\Xi^iy^*$. Clearly, $\Delta z_{ij}^a(0)=0$ and $\Delta z_{i}^a(0)=0$ are stealthy. Similar to (\ref{appendix1_equation3}), one has $\Delta\hat{x}_i^a(1)=\varepsilon \eta(0)\tilde{\Xi}^ix^*$ and $\Delta\hat{x}_h^a(1)=0, \forall h\neq i$. We classify the out-neighbors of the $i$th sensor into $u$ and $s$, where $\gamma_{ui}=1, \gamma_{si}=0, \forall u,s\in \overline{\mathcal{N}}_i$. At time $k=1$, the channels $(i,j)$, $(u,i)$, $(s,t)$ are attacked with the strategies $a_{ij}(1)=\eta(1)\Xi^iy^*-(1-\varepsilon d_i)\eta(0)\tilde{\Xi}^ix^*$, and $a_{ui}(1)=a_{st}(1)=-\varepsilon\eta(0)\tilde{\Xi}^ix^*$, respectively. Note that when $\gamma_{si}=0$, $\mathrm{null}(\tilde{C}_i^a)\subset \mathrm{null}(C_s)$ such that $C_s\tilde{\Xi}^i=0$. Hence, according to $\Delta\hat{x}_h^a(1)=0, \forall h\neq i$, $C_i\Xi^i=0$, $C_i\tilde{\Xi}^i=0$, and $C_s\tilde{\Xi}^i=0$, one can verify that all residues are strictly stealthy. Similar to (\ref{appendix1_equation4}), we can derive that $\Delta\hat{x}_i^a(2)=\varepsilon \eta(1)\tilde{\Xi}^ix^*$. In addition, for the $s$th sensor, $\Delta\hat{x}_s^a(2)=0$ since $a_{st}(1)$ is the opposite of $\Delta\hat{x}_i^a(1)$. The other sensors $\forall h\neq i,s$ do not receive $\Delta\hat{x}_i^a(1)$, and thus $\Delta\hat{x}_h^a(2)=0$. By adopting the similar strategy at time $k=1$, the attacker can keep stealthy and diverge $\Delta\hat{x}_i^a(k)$ with $\eta(k)\to\infty$. The proof is thus completed.
 \hfill\rule{2mm}{2mm}
\end{lemma}

In Lemma \ref{lemma 2}, its second condition is similar to the one in Theorem \ref{theorem 1}. Since $\mathrm{null}(\tilde{C}_i^a)\subset \mathrm{null}(C_i)$, the former can be regarded as a sufficient condition of the latter. In other words, compared with Scenario I, the adversary needs to satisfy a stricter constraint in this case. In addition, as shown in the proof of Lemma \ref{lemma 2}, if $\exists t \in \mathcal{N}_s$, $\gamma_{st}=1$ for $\forall s\in\overline{\mathcal{N}}_i, \gamma_{si}=0$, the attacker can directly manipulate the infinite component of $\Delta\hat{x}_s^a(k)$, such that its evolution is not governed by $\Delta\hat{x}_i^a(k)$. Then, for the $i$th sensor, (\ref{theorem2_equation1}) can be decoupled into the simple form in Lemma \ref{lemma 2}.

In proving the sufficiency of Lemma \ref{lemma 2}, a feasible strategy is provided for the attacker to tamper with partial channels. In this attack case, the adversary only requires to tamper with $1+\bar{d}_i$ channels, where $\bar{d}_i$ is the number of out-neighbors of the $i$th sensor. Hence, the minimum number of attacked channels must not exceed $1+\bar{d}_i$. Besides, it is worth emphasizing that for distributed sensor networks, the heterogeneity of measurement matrixes $C_i$ helps to improve the security. Specifically, for the $i$th sensor, if $\mathrm{rank}([(C_i)^T,(C_s)^T]^T)=n, \forall s \in \overline{\mathcal{N}}_i$, then the attacker can diverge $\Delta\hat{x}_i^a(k)$ only if it attacks all the channels $(s,i)$, which limits the attacker and increases its attack cost. Finally, it should be pointed out that attacks satisfying Definition \ref{definitionl} can maintain strict stealthiness under any residual statistics-based detector including the $\chi^2$ detector (\ref{detector}). In this sense, the system vulnerabilities derived in this section exhibit generality.

\section{Protection Strategy}\label{Section4}
Based on the Euclidean distance between local estimates and channel coding, this section proposes two defense methods to address the vulnerabilities of distributed filtering in Section \ref{Section3}. Moreover, to save coding costs, an algorithm is provided to select critical channels for encoding.

\subsection{Protection strategy based on Euclidean distance between local estimates}\label{Section4.1}
We can first check whether there exists a sensor $i \in\mathcal{V}$, whose parameters satisfy the conditions in Theorem \ref{theorem 1}. If not, the distributed estimator (\ref{DKF}) itself is secure, when the bounded estimation error in Lemma \ref{lemma 1} is tolerable. Otherwise, similar to \cite{zhou2022security}, we can adopt the $\chi^2$ detector to measure the Euclidean distance between $\hat{x}_i(k)$ and $\hat{x}_j(k)$:
\begin{align}
\sum_{s=k-J_{ij}+1}^k [\mu_{ij}(s)]^T(\Sigma_{ij}^x)^{-1}\mu_{ij}(s) \mathop{\stackrel{H_0}\lessgtr}_{H_1} \zeta_{ij},\label{detector1}
\end{align}
where $\mu_{ij}(s)\triangleq\hat{x}_i(s)-\hat{x}_j(s)$ and $\Sigma_{ij}^x=\bar{P}_i+\bar{P}_j-\bar{P}_{ij}-\bar{P}_{ji}$. It is because that according to $\mu_{ij}(k)=(\theta_i-\theta_j)\otimes I_ne(k)$ and $e(k)\sim\mathcal{N}(0,\bar{P})$, $\mu_{ij}(k)$ follows the zero-mean Gaussian distribution with covariance $\Sigma_{ij}^x$. Note that the defense method in \cite{zhou2022security} fundamentally relies on the stochastic $\chi^2$ detection \cite{li2016stochastic}, which is an alternative to (\ref{detector1}). Define $\mu_{ij}^a(k)=\hat{x}_i^a(k)-\hat{x}_j^a(k)-\gamma_{ij}a_{ij}(k)$ and $\Delta\mu_{ij}^a(k)=\mu_{ij}(k)-\mu_{ij}^a(k)$. When (\ref{detector1}) or the one in \cite{zhou2022security} is employed, the attack is still strictly stealthy only if $\mu_{ij}^a(k)\sim\mathcal{N}(0,\Sigma_{ij}^x)$. Thus, compared with the detection based on $z_{ij}(k)$, (\ref{detector1}) is not subject to the measurement matrix $C_i$. Similar to Theorem \ref{theorem 1}, we first consider the special case of $\gamma_{ij}=1, \forall j \in \mathcal{N}_i, i \in\mathcal{V}$, and reveal the necessary and sufficient condition required by the attacker.
\begin{theorem} \label{theorem 3}
Under Assumptions \ref{assumption1}-\ref{assumption3}, for the $i$th sensor, the attack (\ref{attack model}) can diverge $\Delta\hat{x}_i^a(k)$ and keep strictly stealthy under detectors (\ref{detector}) and (\ref{detector1}), if and only if 1) $\mathrm{rank}(C_i)<n$, 2) the matrix $A$ is unstable, and 3) $\exists x_0\in \mathbb{R}^{l_i}\neq 0$, $\Xi^ix_k=A^k\Xi^ix_0$ has a nontrivial solution $x_k$ for $\forall k$.

\noindent\textbf{Proof.} \rm The necessity is proved at first. By substituting $\Delta\hat{x}_j^a(k)+\gamma_{ij}a_{ij}(k)-\Delta\hat{x}_i^a(k)=\Delta\mu_{ij}^a(k)$ and $\Delta\hat{x}_i^a(k)=\Xi^i \alpha^i(k)+\rho^i(k)$ into (\ref{DKF_attack_difference}), we have
\begin{align*}
 \Xi^i \alpha^i(k+&1)=A\Xi^i \alpha^i(k)+(A-\bar{K}_iC_i)\rho^i(k)-\rho^i(k+1)\nonumber\\
                    &+\varepsilon A\sum\limits_{j\in \mathcal{N}_i}[\Delta\mu_{ij}^a(k)]\triangleq A\Xi^i \alpha^i(k)+\tilde{\mu}^i(k),
\end{align*}
where $\tilde{\mu}^i(k)$ is also a zero-mean variable with bounded $\mathbb{E}[\Vert\tilde{\mu}^i(k)\Vert_2]$. Define $\eta_{k}\triangleq \Vert\alpha^i(k)\Vert_2$. When $\eta_{k}\to\infty$ as time goes by, one has $\lim_{k\to\infty}\{\eta_{k+1}^{-1}[\Xi^i\alpha^i(k+1)-A\Xi^i \alpha^i(k)]\}=0$. Then, we can deduce that $\lim_{k\to\infty}\{\eta_{k+l}^{-1}[\Xi^i\alpha^i(k+l)-A^{l}\Xi^i \alpha^i(k)]\}=0$, which means $\exists x_0\in \mathbb{R}^{l_i}\neq 0$, $\Xi^ix_k=A^k\Xi^ix_0$ has a nontrivial solution $x_k$ for $\forall k$. Define $\varpi_{min}^i$ and $\varpi_{max}^i$ as the minimum and maximum eigenvalues of $[\Xi^i]^T\Xi^i$, respectively. If $A$ is stable, one can obtain that
\begin{align*}
 \varpi_{min}^i\leq&\eta_{k+l}^{-1}\Vert\Xi^i\alpha^i(k+l)\Vert_2=\eta_{k+l}^{-1}\Vert A^{l}\Xi^i \alpha^i(k)\Vert_2\nonumber\\
                \leq&\eta_{k+l}^{-1}\Vert A^{l}\Vert_2 \varpi_{max}^i\eta_{k}\leq\eta_{k+l}^{-1}[\iota_A(\kappa_A)^l] \varpi_{max}^i\eta_{k},
 \end{align*}
where $\iota_A$ and $0\leq\kappa_A<1$ are constants such that $\Vert A^s\Vert_2\leq \iota_A(\kappa_A)^s$ if $A$ is stable \cite{boem2017distributed}. When $\eta_{k}$ is fixed, the above inequality holds, only if $\eta_{k+l}$ decreases monotonically with the increase of $l$. Thus, it contradicts the attack object $\eta_{k+l}\to\infty$. By contradiction, it implies that $A$ is unstable.

Finally, we prove the sufficiency. At time $k=0$, the attack is designed as $a_{ij}(0)=-\Delta\hat{x}_j^a(0)+\Delta\hat{x}_i^a(0)+\eta(0)\Xi^ix_0$, where $\eta(0)\in \mathbb{R}$ is a small scalar. Then, $\Delta z_{ij}^a(0)=\Delta z_{i}^a(0)=0$, and $\Delta\mu_{ij}^a(0)=\eta(0)\Xi^ix_0$ are almost strictly stealthy. In addition, through the iteration of (\ref{DKF_attack_difference}), one has $\Delta\hat{x}_i^a(1)=\varepsilon d_i\eta(0)\Xi^ix_1$ when $\forall \gamma_{ij}=1$. At time $k=1$, $a_{ij}(1)=-\Delta\hat{x}_j^a(1)+\Delta\hat{x}_i^a(1)$. Then, $\Delta z_{ij}^a(1)=\Delta z_{i}^a(1)=\Delta\mu_{ij}^a(1)=0$ are strictly stealthy. Besides, one has $\Delta\hat{x}_i^a(2)=\varepsilon d_i\eta(0)\Xi^ix_2$. At time $k=2,...$, $a_{ij}(k)$ is designed to be the same as the one at time $k=1$. Consequently, $\Delta z_{i}^a(k)$, $\Delta z_{ij}^a(k)$, and $\Delta\mu_{ij}^a(k)$ can bypass the detection, and $\Delta\hat{x}_i^a(k)=\varepsilon d_i\eta(0)\Xi^ix_k$. Since $A$ is unstable, and $A^k\Xi^ix_0\neq 0, \forall k$, it implies that $\Vert x_k\Vert_2\to\infty$ as time goes to infinity. The proof is thus completed.
 \hfill\rule{2mm}{2mm}
\end{theorem}

In Theorem \ref{theorem 3}, its first condition guarantees that $C_i$ possesses a non-empty null space, such that there exist infinite components in $e_i^a(k)$ which remain stealthy under the detector (\ref{detector}); its second condition enables stealthy attacks to diverge $e_i^a(k)$ under the constraint of the detector (\ref{detector1}); its third condition ensures that with the evolution of iterative formula (\ref{DKF_attack}), the infinite component in $e_i^a(k)$ lies within the null space of $C_i$ for all times $k$, thereby anomalies in $e_i^a(k)$ remain persistently undetectable. Theorem \ref{theorem 3} illustrates that the detection (\ref{detector1}) based on $\mu_{ij}(k)$ cannot fully guarantee the security of the distributed estimator (\ref{DKF}). Nevertheless, since Theorem \ref{theorem 3} requires more and stricter conditions than Theorem \ref{theorem 1}, this method can still limit attackers and mitigate security risks to a certain extent. For instance, compared with Theorem \ref{theorem 1}, Theorem \ref{theorem 3} further requires that $A$ is unstable and $\Xi^i\in \mathbb{R}^{n}$ is also an unstable eigenvector of $A$ when $\mathrm{rank}(C_i)=n-1$. In addition, as shown in the proof of Theorem \ref{theorem 1}, the attacker can even diverge $\Delta\hat{x}_i^a(k)$ at the initial time $k$ in the original system. However, from the attack case corresponding to Theorem \ref{theorem 3}, we can see that the divergence of $\Delta\hat{x}_i^a(k)$ relies on the instability of $A$, such that the attacker needs to accumulate its attack effect over time.

\begin{remark}\label{remark6}
\rm If $\Xi^i$ is also an eigenvector of $A$, it implies that the observability matrix $\Omega_i\triangleq[C_i^T,(C_iA)^T,...,(C_iA^n)^T]^T$ is not full column rank, since the equation $\Omega_ix=0$ has the nontrivial solution $x=\Xi^i$. In other words, $(A,C_i)$ must not be locally observable in this case. Note that the sensor network is only required to satisfy the detectability condition in Assumption \ref{assumption3}. Therefore, the above vulnerability may indeed exist in some nodes of sensor networks.
\end{remark}

Next, we explore the general case that the attacker invades a part of channels. When there is no attack on the channel $(i,j), j \in \mathcal{N}_i$, the transmitted data $\Delta\hat{x}_j^a(k)$ can deceive the detector (\ref{detector1}), only if $\Delta\hat{x}_j^a(k)$ and $\Delta\hat{x}_i^a(k)$ have the same infinite components, which depend on $\Xi^j$ and $\Xi^i$, respectively. That is, $\Xi^j\alpha^j(k)=\Xi^i\alpha^i(k)$. Similarly, if the channels $(s,i), s \in \overline{\mathcal{N}}_i$ and $(s,t), t\neq i, s \in \bar{\mathcal{N}}_t$, are not attacked, it means that $\Xi^i\alpha^i(k)=\Xi^s\alpha^s(k)=\Xi^t\alpha^t(k)$. In a word, $\Vert\Delta\hat{x}_i^a(k)\Vert_2\to\infty$ is determined by all the sensors including $i$, $\gamma_{ij}=0, \forall j \in \mathcal{N}_i$, $\gamma_{si}=0, \forall s \in \overline{\mathcal{N}}_i$, $\gamma_{st}=0, t\neq i, \forall s \in \overline{\mathcal{N}}_t$, and so on. Note that the constraint of $\Xi^t$ on $\Vert\Delta\hat{x}_i^a(k)\Vert_2\to\infty$ is not related to the directionality of the channel $(s,t)$. In view of this, we first transform $A_{L}-A_{\gamma}$ from the directed graph into the undirected graph, which is defined as $\tilde{A}_{(L,\gamma)}$. Then, we can calculate the reachability matrix $\tilde{R}_{(L,\gamma)}$ corresponding to $\tilde{A}_{(L,\gamma)}$. Based on the $i$th row (or column) of $\tilde{R}_{(L,\gamma)}$, we can select the relevant measurement matrixes, and stack them into a column $\check{C}_i^a$. In summary, the infinite component of  $\Vert\Delta\hat{x}_i^a(k)\Vert_2$ is governed by $\mathrm{null}(\check{C}_i^a)$. Clearly, $\tilde{R}_{(L,\gamma)}$ covers the nonzero elements of $A_{L}-A_{\gamma}$, which indicates that $\mathrm{null}(\check{C}_i^a) \subset \mathrm{null}(\tilde{C}_i^a)$. Thus, compared with the original detection system, the attack space is further limited by the detector (\ref{detector1}). The insecurity of distributed estimation in this case is analyzed below.

\begin{lemma} \label{lemma 3}
Under Assumptions \ref{assumption1}-\ref{assumption3}, when partial channels are attacked, the attack (\ref{attack model}) can diverge $\Delta\hat{x}_i^a(k)$ and bypass the detection of $\Delta z_{i}^a(k)$, $\Delta z_{ij}^a(k)$, and $\Delta\mu_{ij}^a(k)$, only if 1) $\mathrm{rank}(\check{C}_i^a)<n$, 2) the matrix $A$ is unstable, and 3) $\exists x_0\in \mathbb{R}^{\check{l}_i}\neq 0$, $\check{\Xi}^ix_k=A^k\check{\Xi}^ix_0$ has a nontrivial solution $x_k$ for $\forall k$, where $\check{\Xi}^i$ and $\check{l}_i$ correspond to $\check{C}_i^a$, and are similar to the case of $\tilde{C}_i^a$.

\noindent\textbf{Proof.} The proof is similar to Theorem \ref{theorem 3}, and thus is omitted here.
 \hfill\rule{2mm}{2mm}
\end{lemma}

\subsection{Protection strategy based on coding scheme}\label{Section4.2}
\begin{figure}[!tbp]
\begin{minipage}[t]{1\linewidth}
      \centering
      \subfigure[]{\includegraphics[scale=0.48]{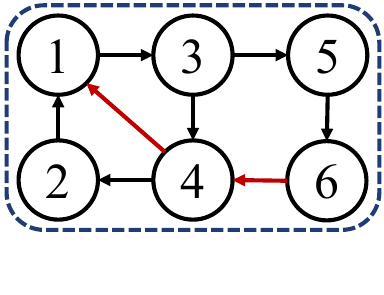}}
      \subfigure[]{\includegraphics[scale=0.49]{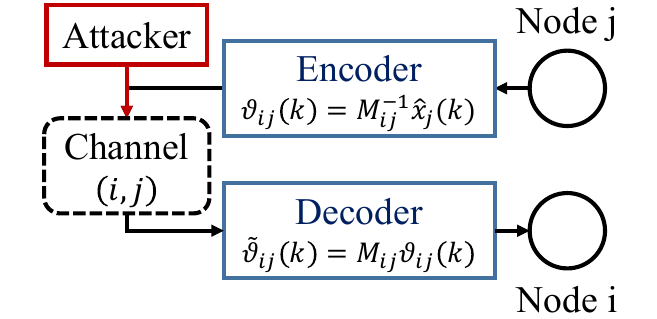}}
      \caption{System diagram: (a) red lines denote the encoded channels, (b) the channel $(i,j)$ is protected by the coding scheme.}
      \label{Fig_systemblock2}
\end{minipage}
\end{figure}

Guided by the security analysis in Section \ref{Section3}, this section aims to develop targeted defense mechanisms to comprehensively address the security vulnerabilities in Definition \ref{definitionl}, thereby effectively safeguarding the distributed estimation (\ref{DKF}) against stealthy attacks (\ref{attack model}). When at least one condition in Theorem \ref{theorem 3} is not satisfied for each sensor, the detector (\ref{detector1}) is sufficient to protect the security of distributed filtering (\ref{DKF}). Otherwise, we can adopt the coding scheme \cite{miao2016coding,zhou2022watermarking} to compensate for the vulnerabilities left by (\ref{detector1}). Different from (\ref{detector1}), the coding scheme involves encryption and decryption of communication data, and its effectiveness depends entirely on its confidentiality. That is, it may be at risk of being cracked by the attackers. Thus, if the conditions in Theorem \ref{theorem 3} are not satisfied, the selection priority of (\ref{detector1}) should be higher than the coding scheme. In a word, the coding scheme serves as a complement to (\ref{detector1}).

Note that the sensor network is constrained by the limited energy of onboard batteries, while the coding scheme needs to consume a certain amount of coding costs to generate feasible coding matrixes. Besides, the overall coding cost of the entire sensor network increases monotonically with the increase in the number of coding channels. Hence, encode all channels is a high requirement especially in large-scale sensor networks. As seen in Fig. \ref{Fig_systemblock2}, we consider a general framework of partial channel encoding. If the channel $(i,j)$ is selected for encoding, the $j$th sensor will encode $\hat{x}_j^a(k)$ based on an invertible matrix $M_{ij}^{-1}\in \mathbb{R}^{n\times n}$ (the time-varying version will be introduced later). Hence, instead of $\hat{x}_j^a(k)$, the information transmitted to the neighbor sensor $i$ is the encoded data $\vartheta_{ij}^a(k)$, which is
\begin{align}
\vartheta_{ij}(k)=M_{ij}^{-1}\hat{x}_j^a(k).\label{encoder}
\end{align}
For the data received from the channel $(i,j)$, the $i$th sensor can utilize $M_{ij}$ to decode it as:
\begin{align}
\hat{x}_{ij}^e(k)=M_{ij}\tilde{\vartheta}_{ij}(k),\label{decoder}
\end{align}
where $\tilde{\vartheta}_{ij}(k)=\vartheta_{ij}(k)+\gamma_{ij}a_{ij}(k)$. If the channel $(i,j)$ is not attacked, i.e., $\gamma_{ij}=0$, the decoded data $\hat{x}_{ij}^e(k)$ can be reverted to $\hat{x}_j^a(k)$. Otherwise, $\hat{x}_{ij}^e(k)=\hat{x}_j^a(k)+M_{ij}a_{ij}(k)$, whose induced residues are affected by $M_{ij}$, thereby triggering the alarm. If the attacker is not aware of the coding scheme, it still adopts the original attack sequence $\{a_{ij}(k)\}$ in Section \ref{Section3}. In this case, the following lemma provides a sufficient condition to design $M_{ij}$, which guarantees that any $\{a_{ij}(k)\}$ loses its stealthiness under the detector (\ref{detector}).

\begin{lemma} \label{lemma 4}
For the original attack $\{a_{ij}(k)\}$ that can deceive the detector (\ref{detector}) and diverge $\Delta\hat{x}^a(k)$, if the channel $(i,j)$ is attacked, and there exists a coding matrix $M_{ij}$ such that $\forall x_1\in \mathbb{R}^{n}\neq 0$, $\mathrm{rank}(\Theta)<\mathrm{rank}([\Theta,x_2])$, where
\begin{align}
\Theta=\left(
  \begin{array}{ccccc}
      \Xi^i  & -\Xi^j & 0\\
      M_{ij}\Xi^i  & (I_n-M_{ij})\Xi^j &-\Xi^i \\
  \end{array}
\right),\label{lemma 4_equation1}
\end{align}
and $x_2=[(x_1)^T,0^T]\in \mathbb{R}^{2n}$, then the residue $\Vert\Delta z_{ij}^a(k)\Vert_2\to\infty$ when $\Vert a_{ij}(k)\Vert_2\to\infty$.

\noindent\textbf{Proof.} The proof is shown in Appendix \ref{appendix6}.
\hfill\rule{2mm}{2mm}
\end{lemma}

For the system configured with both the detectors (\ref{detector}) and (\ref{detector1}), a sufficient condition to design $M_{ij}$ is as follows.

\begin{theorem} \label{theorem 4}
For the attack sequence $\{a_{ij}(k)\}$ that can bypass both (\ref{detector}) and (\ref{detector1}) and diverge the estimation error $\Delta\hat{x}^a(k)$, if the channel $(i,j)$ is attacked and there exists a coding matrix $M_{ij}$ such that $\mathrm{rank}(I_n-M_{ij})=n$, then the residue $\Vert\Delta\mu_{ij}^a(k)\Vert_2\to\infty$ when $\Vert a_{ij}(k)\Vert_2\to\infty$.

\noindent\textbf{Proof.} The proof is similar to Lemma \ref{lemma 4}, and thus is omitted here.
 \hfill\rule{2mm}{2mm}
\end{theorem}

Compared with Lemma \ref{lemma 4}, Theorem \ref{theorem 4} requires a weaker constraint on the feasible matrix $M_{ij}$, since there must exist $M_{ij}$ such that $I_n-M_{ij}$ is full rank. Hence, the joint protection based on the detector (\ref{detector1}) and coding scheme can relax the design requirements of $M_{ij}$. Besides, it can reduce the computation cost to obtain $M_{ij}$, which only involves solving the matrix rank with the computational complexity $O(n^3)$. Note that as stated in \cite{miao2016coding}, compared with the encryption that requires highly complex nonlinear operations, the coding scheme is a low-cost alternative at the expense of a certain degree of data confidentiality, and is more suitable for wireless sensor networks.

When the attacker is aware of the encoding scheme, it may learn the knowledge of $M_{ij}$ by eavesdropping the transmitted data. If the exact value of $M_{ij}$ is known, the attacker can redesign the attack sequence to circumvent the coding scheme. Based on the communication protocol of distributed filtering (\ref{DKF}), the information broadcasted from the $i$th sensor to its out-neighbors is the same, i.e., $\hat{x}_i^a(k)$. Hence, as depicted in Fig. \ref{Fig_systemblock2}, when only partial channels are encoded, the attacker can intercept both the output and input of (\ref{encoder}) from the channel $(i,j)$ and the unprotected channel $(t,j)$, $t\in \bar{\mathcal{N}}_j, t\neq i$, respectively.

We consider the worst case that the attacker can distinguish the encoded channels. Then, without any prior knowledge of $M_{ij}$, the attacker can acquire $n$ equations with respect to $n^2$ variables to estimate $M_{ij}$. By observing several time steps, the attacker can obtain a series of information sets $\{\hat{x}_j^a(s),\vartheta_{ij}(s)\}$, where $s=k,...,k+l_o$, and $l_o$ denotes the duration. Define $\Upsilon\triangleq [\hat{x}_j^a(k),...,\hat{x}_j^a(k+l_o)]$ and $\Gamma\triangleq [\vartheta_{ij}(k),...,\vartheta_{ij}(k+l_o)]$. Then, for the $s$th row of $M_{ij}^{-1}$, the attacker can construct a non-homogeneous linear equation $\Upsilon^T([M_{ij}^{-1}]^{[s]})^T=(\Gamma^{[s]})^T$, which has a unique solution only if $\mathrm{rank}(\Upsilon)=n$. In other words, the attacker needs at least $l_o\geq n$ observations to calculate $M_{ij}^{-1}$. It suggests that the coding matrix should be time-varying with the maximum dwell time less than $n$.

As pointed out in \cite{miao2016coding}, for the coding scheme in networked control systems, the learning of $M_{ij}$ involves solving bilinear equation with noise. Hence, compared with \cite{miao2016coding}, it is more challenging to protect the privacy of $M_{ij}$ in distributed sensor networks. Moreover, if all channels are encoded, the exact value of the input of (\ref{encoder}) is unavailable to the attacker. It indicates that encoding partial channels can reduce the costs at the expense of sacrificing some confidentiality of $M_{ij}$.
\begin{remark}\label{remark7}
\rm In \cite{zhou2022watermarking}, the coding matrix was designed as $M_{ij}(k)=\lambda I_n+\beta(k)\beta(k)^T$, where $\lambda>0$ is a scalar, and $\beta(k)$ is a zero-mean Gaussian variable with covariance $\Sigma_{\beta}$. Besides, in \cite{zhou2022watermarking}, $\Vert\Delta z_{ij}^a(k)\Vert_2\to\infty$ can be ensured by choosing $\varrho_{ij}(k)\to\infty$, where $\varrho_{ij}(k)$ is the $2$-norm of $M_{ij}(k)$. However, when partial channels are encoded, the attacker can obtain some prior information of $\varrho_{ij}(k)$ even based on the one-step observation. Specifically, by utilizing the rayleigh quotient, the attacker has $\delta_{min}^{ij}\leq [(\vartheta_{ij}(k))^T\vartheta_{ij}(k)]/[(\hat{x}_j^a(k))^T(\hat{x}_j^a(k))]$, where $\delta_{min}^{ij}$ is the minimum singular value of $M_{ij}^{-1}(k)$. Note that $\delta_{min}^{ij}=(\Vert M_{ij}(k)\Vert_2)^{-1}=(\varrho_{ij}(k))^{-1}$, such that a lower bound $\tilde{\varrho}_{ij}(k)$ of $\varrho_{ij}(k)$ is exposed to the attacker. In this case, $\Vert\Delta z_{ij}^a(k)\Vert_2$ may not tend to infinity if the attacker adjusts $a_{ij}(k)$ based on $\tilde{\varrho}_{ij}(k)$. Hence, different from \cite{zhou2022watermarking}, we do not require $M_{ij}(k)$ to satisfy $\varrho_{ij}(k)\to\infty$.
\end{remark}

Similar to the coding scheme in \cite{zhou2022watermarking,miao2016coding} and the moving target defense in \cite{griffioen2020moving}, we adopt the cryptographically secure pseudo random number generator (PRNG) to update the time-varying coding matrix $M_{ij}(k)$. For the channel $(i,j)$, its sending and receiving sides hold an identical generator seed, such that $M_{ij}(k)$ is synchronized on both sides. Besides, the seed is analogous to a secret key, and should be hidden from the adversary. In a word, $M_{ij}(k)$ is deterministic to the defender, while is unknown and random to the attacker. Since the seed of each channel can be different from each other, some existing techniques of secret key distribution can be applied to configure seeds for sensor networks. With a high switching frequency of $M_{ij}(k)$, the adversary cannot ensure that the attack sequence still remains stealthy before $M_{ij}(k)$ is cracked. Similar to \cite{miao2016coding}, we consider that the attacker calculates an estimated coding matrix $\tilde{M}_{ij}(k)$ and redesigns the injected data as $a_{ij}^*(k)=\tilde{M}_{ij}^{-1}(k)a_{ij}(k)$, where $a_{ij}(k)$ is the original one. In the following, we show that the time-varying coding scheme can effectively protect the channel $(i,j)$ by assisting the detector (\ref{detector}) against $a_{ij}^*(k)$.

\begin{theorem} \label{theorem 5}
For the case that the encoded channel $(i,j)$ is injected by the redesigned sequence $\{a_{ij}^*(k)\}$, if $\tilde{M}_{ij}(k)\neq M_{ij}(k)$, then the attacker (\ref{attack model}) cannot guarantee that the residue $\Delta z_{ij}^a(k)$ is strictly stealthy under the detector (\ref{detector}).

\noindent\textbf{Proof.} The proof is shown in Appendix \ref{appendix7}.
 \hfill\rule{2mm}{2mm}
\end{theorem}

Similarly, we can further extend the above result to the case that both the detectors (\ref{detector}) and (\ref{detector1}) are employed. Hence, it demonstrates that the coding scheme described by (\ref{encoder}) and (\ref{decoder})  serves as a low-cost yet sufficiently effective countermeasure to address the insecurity of distributed filtering (\ref{DKF}) in Definition \ref{definitionl}.

\subsection{Allocation strategy of encoded channels}\label{Section4.3}

The coding scheme can prevent the encoded channels from being attacked. However, the adversary may identify the encoded channels and only attack the unprotected channels based on the strategy in Section \ref{Section3.2}. In view of this, we can restrict the attack by properly allocating encoded channels. Define a binary variable $\lambda_{si}=0$ or $1$ to represent whether the channel $(s,i)$ is encoded, and stack $C_s, \lambda_{si}=1, \forall s \in \overline{\mathcal{N}}_i$ and $C_i$ into a column, i.e., $\bar{C}_i\triangleq [...,(C_s)^T...,(C_i)^T]^T$. Then, if $\forall i \in\mathcal{V}$, $\mathrm{rank}(\bar{C}_i)=n$ or the second one in Lemma \ref{lemma 2} does not hold, the stealthy attack that intrudes unprotected channels can only yield bounded estimation error. To reduce the resource consumption of the coding scheme, we needs to minimize the number of encoded channels under the premise of satisfying the above conditions. Thus, the allocation of encoded channels can be formulated as an optimization problem, subject to the constraint of satisfying the first condition.
\begin{align}
&\min \quad \sum\nolimits_{s=1}^{N}\lambda_{si} \nonumber\\
& \begin{array}{r@{\quad}l@{}l@{\quad}l}
s.t.& \mathrm{rank}(\bar{C}_i)=n. \label{min 1}
\end{array}
\end{align}
For the $i$th sensor, the optimization space of (\ref{min 1}) is the combination of $C_s, \forall s \in \overline{\mathcal{N}}_i$, whose cardinality is $2^{\bar{d}_i}$. Besides, (\ref{min 1}) suggests to encode the channel $(s,i)$, where $C_s$ and $C_i$ are highly heterogeneous. If $\exists s \in \overline{\mathcal{N}}_i, \mathrm{rank}([(C_i)^T,(C_s)^T]^T)=n$, it is sufficient to only encode one channel $(s,i)$ to ensure the security of the $i$th sensor. Corresponding to $\bar{C}_i$, we rewrite $\tilde{\Xi}^i$ in Lemma \ref{lemma 2} as $\bar{\Xi}^i$. Then, we can further construct the following optimization problem for the second condition.
\begin{align}
&\min \quad \sum\nolimits_{s=1}^{N}\lambda_{si} \nonumber\\
& \begin{array}{r@{\quad}l@{}l@{\quad}l}
s.t.& \mathrm{rank}(\bar{C}_i)<n,\\
    &\forall x\neq 0, \mathrm{rank}(A\Xi^i)<\mathrm{rank}([A\Xi^i,\bar{\Xi}^ix]),\label{min 2}
\end{array}
\end{align}
where the first constraint is a prerequisite for the second one, i.e., $\bar{\Xi}^i\neq 0$ when $\mathrm{rank}(\bar{C}_i)<n$. Note that (\ref{min 1}) and (\ref{min 2}) have different feasible optimization regions. Hence, the allocation strategy corresponds to the minimum between the optimal solutions of (\ref{min 1}) and (\ref{min 2}). It should be emphasized that (\ref{min 1}) and (\ref{min 2}) minimize the number of encoded channels from the local perspective of each sensor. However, if $\exists i\in\mathcal{V}$, neither (\ref{min 1}) nor (\ref{min 2}) has a feasible solution satisfying the constraints, Lemma \ref{lemma 2} cannot ensure the security of this sensor even if all channels $(s,i)$ are encoded. In this case, we need to utilize Theorem \ref{theorem 2} to allocate encoded channels from the global perspective of sensor networks, which may consume more computing resources. In summary, we provide a procedure for selecting encoded channels to save coding costs while ensuring the detection capability of the detector (\ref{detector}).
\begin{algorithm}
\caption{Strategy for configuring encoded channels}
\begin{algorithmic}[1] \label{algorithm 1}
\REQUIRE Measurement matrixes $C_i, \forall i \in\mathcal{V}$, Laplacian matrix $L$, and system matrix $A$.
\FOR{$i=1$ \TO $N$}
\IF {$\mathrm{rank}(C_i)<n$}
\STATE  Calculate $\Xi^i$ and $l_i$ based on $\mathrm{null}(C_i)$.
\IF {$\exists x\neq 0, \mathrm{rank}(A\Xi^i)=\mathrm{rank}([A\Xi^i,\Xi^ix])$}
\STATE Calculate (\ref{min 1}) and (\ref{min 2}), and return the solution with the minimum cardinality. If both (\ref{min 1}) and (\ref{min 2}) are unsolvable, then interrupt the algorithm.
\ENDIF
\ENDIF
\ENDFOR
\end{algorithmic}
\end{algorithm}

Recall that the $i$th sensor itself is secure when the conditions in Theorem \ref{theorem 1} do not hold. Hence, Steps 2 and 4 can avoid configuring unnecessary encoded channels for these sensors. Besides, many existing methods such as greedy algorithm can be applied to solve the combinatorial optimization problems (\ref{min 1}) and (\ref{min 2}) in Step 5. It is well known that the singular value decomposition (SVD) can be utilized to calculate the rank and null space of a matrix $X\in \mathbb{R}^{m\times n}$, with a computational complexity of $O(min(mn^2,m^2n))$ \cite{strang2022introduction}. Hence, one can deduce that the computational complexity of  Algorithm 1 is $O(2^{\max_i(d_i)}Nn\phi^2)$, where $\phi=\max(n,\max_{i \in\mathcal{V}}(m_i+\sum_{s \in \overline{\mathcal{N}}_i}m_s))$. Finally, for the system protected by both the detectors (\ref{detector}) and (\ref{detector1}), we can further reduce the number of channels that need to be encoded. It is because that compared with Lemma \ref{lemma 2}, the attacker is limited by stricter constraints in Lemma \ref{lemma 3}. Similar to Algorithm \ref{algorithm 1}, one can also develop a corresponding procedure to minimize the number of encoded channels in this case.

\begin{figure*}[t]
	\centering
	\begin{minipage}{0.48\linewidth}
		\centering
		\includegraphics[width=0.9\linewidth]{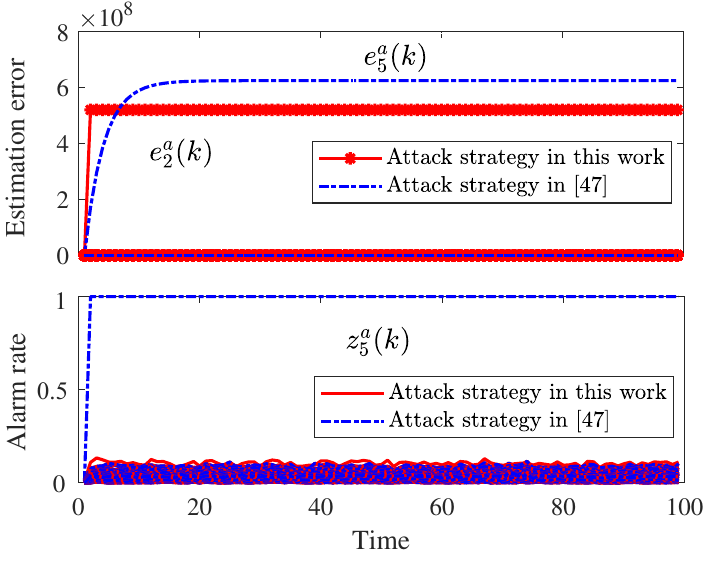}
		\caption{Comparison between the attack strategy satisfying Lemma \ref{lemma 2} and the one in \cite{zhou2022watermarking} in terms of the estimation error of the distributed estimator (\ref{DKF}) and the alarm rate of the detector (\ref{detector}).}
		\label{simulation 1}
	\end{minipage}
    \hspace{5pt}
	\begin{minipage}{0.48\linewidth}
		\centering
		\includegraphics[width=0.9\linewidth]{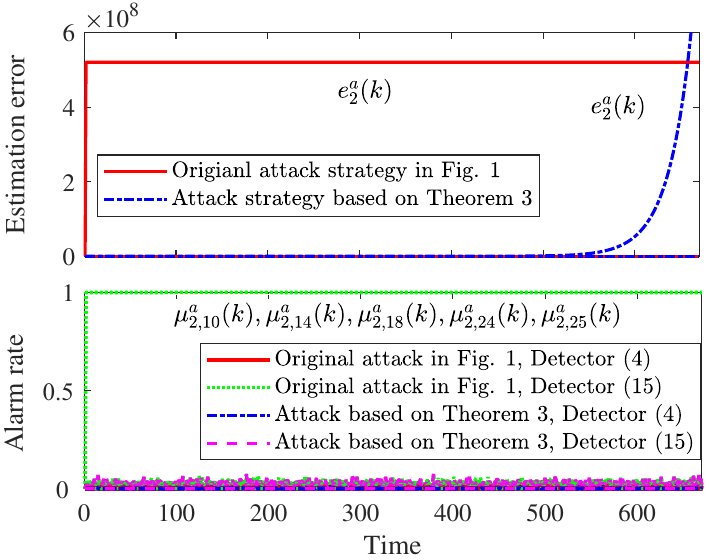}
		\caption{Comparison between the attack strategy satisfying Lemma \ref{lemma 2} and the one satisfying Theorem \ref{theorem 3} in terms of the estimation error of the distributed estimator (\ref{DKF}) and the alarm rate of the detectors (\ref{detector}) and (\ref{detector1}).}
		\label{simulation 2}
	\end{minipage}
	\begin{minipage}{0.48\linewidth}
		\centering
		\includegraphics[width=0.9\linewidth]{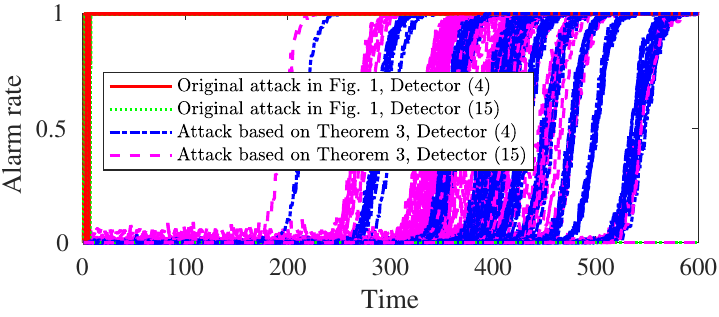}
		\caption{Under the protection of the coding scheme, the alarm rate of the detectors (\ref{detector}) and (\ref{detector1}), when the attacker exploits the original attack strategies.}
		\label{simulation 3}
	\end{minipage}
    \hspace{5pt}
	\begin{minipage}{0.48\linewidth}
		\centering
		\includegraphics[width=0.9\linewidth]{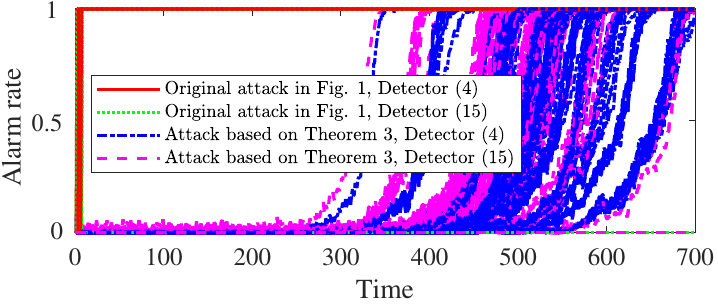}
		\caption{Under the protection of the coding scheme, the alarm rate of the detectors (\ref{detector}) and (\ref{detector1}), when the attack sequence is constructed based on the estimated coding matrix $\tilde{M}_{ij}(k)$.}
		\label{simulation 4}
	\end{minipage}
\end{figure*}

\section{Simulation Examples}\label{Section5}

In this section, a sensor network composed of $N=30$ sensors is considered to monitor the state of the double inverted
pendulum model in \cite{lin2017optimal} with following parameters:
\begin{align*}
A=\left(
    \begin{array}{cccccc}
     1&  -0.0004&  0&  0.0093&  0&  0 \\
     0&  1.0034&  -0.0010&  0.0016&  0.0090&   0.0003\\
     0&  -0.0038&  1.0032&  -0.0004&  0.0008&  0.0094 \\
     0&  -0.0786&  0.0063&  0.8730&  0.0083&   -0.0048\\
     0&  0.6544&  -0.2380&  0.3101&  0.9034&   0.0664\\
     0&  -0.7149&  0.6137&  -0.0751&  0.1579&   0.8770\\
    \end{array}
\right),
\end{align*}
and $Q=0.01I_6$. Besides, the measurement matrix $C_i\in\mathbb{R}^{5\times 6}$ is randomly generated, and the covariance of its measurement noise is set to $R_i=\nu_iI_2$, where $\nu_i\in (0,1]$. For the distributed estimator (\ref{DKF}), its consensus gain is selected as $\varepsilon=0.05$. Moreover, with the window size $J_i=1$ and the confidence coefficient $95\%$, the thresholds of the detectors (\ref{detector}) and (\ref{detector1}) are determined to be $11.07$ and $12.59$, respectively.

For the $2$nd sensor, the null space of $C_2$ consists of $\Xi=[-0.0062,0.1376,-0.1984,-0.0211,0.5748,-0.7816]^T$, which is also an unstable eigenvector of $A$ with the eigenvalue $\lambda_a=1.0405$. That is, the $2$nd sensor satisfies all the conditions in Theorem \ref{theorem 1}. Notice that the $2$nd sensor has a unique out-neighbor, i.e., the $14$th sensor. Hence, based on Lemma \ref{lemma 2}, the adversary aiming to destabilize $e^a_2(k)$ only needs to intrude two channels $(14,2)$ and $(2,10)$. Specifically, at time $k=0$, the channel $(2,10)$ is injected with the false data $\eta \Xi$, where $\eta=10^{10}$ is arbitrarily large. After this, the attack sequences on the channels $(14,2)$ and $(2,10)$ are $-\varepsilon \lambda_a\eta\Xi$ and $\eta \Xi-(1-\varepsilon d_2)\lambda_a\eta\Xi$, respectively. In contrast, the $5$th sensor fully meets the requirements in \cite{zhou2022watermarking}, but does not satisfy the second condition in Theorem \ref{theorem 1}. For comparison, we consider that the adversary adopts Algorithm 1 in \cite{zhou2022watermarking} to generate false data to corrupt $e^a_5(k)$. Via $1000$ Monte Carlo simulations, Fig. \ref{simulation 1} demonstrates the estimation performance and the alarm rate of the detector (\ref{detector}) under the above two different attacks. As can be seen, the proposed attack strategy can not only diverge $e^a_2(k)$, but also avoid both $z^a_{ij}(k)$ and $z^a_{i}(k)$ being detected. In contrast, the attack in \cite{zhou2022watermarking} cannot prevent the alarm rate of $z^a_{i}(k)$ from rising to $100\%$.

Fig. \ref{simulation 2} first evaluates the stealthiness of the attack in Fig. \ref{simulation 1}, when the detector (\ref{detector1}) is further employed. It shows that $\mu_{ij}(k)$ under such an attack can trigger the alarm with the probability $100\%$, which means that the detector (\ref{detector1}) can improve the detection capability. However, since all the conditions of Theorem \ref{theorem 3} still hold for the $2$nd sensor, the detector (\ref{detector1}) cannot fully overcome the vulnerability. Following the strategy in Theorem \ref{theorem 3}, the adversary attacks the channels $(2,10)$, $(2,14)$, $(2,18)$, $(2,24)$, $(2,25)$, and $(14,2)$. At time $k=0$, the first five channels are injected with $\tilde{\eta} \Xi$, where $\tilde{\eta}=0.01$ is chosen to be small. For the other time $k$, the attack sequences on the first five channels are $\varepsilon d_2\tilde{\eta}(\lambda_a)^{k-1}\Xi$, while the ones on the last channel are the opposite. Fig. \ref{simulation 2} shows that the redesigned attack strategy can deceive both the detectors (\ref{detector}) and (\ref{detector1}).

Based on Algorithm \ref{algorithm 1}, the other insecure sensors include the $20$th and $27$th sensors. Then, only three channels $(14,2)$, $(19,20)$, and $(17,27)$ need to be encoded. According to whether the system is configured with the detector (\ref{detector1}), $M_{ij}(k)$ should be generated based on Lemma \ref{lemma 4} and Theorem \ref{theorem 4}, respectively. Fig. \ref{simulation 3} shows that with the help of the coding scheme, the detectors (\ref{detector}) and (\ref{detector1}) can effectively detect the originally undetectable attacks. Notice that the alarm rate for the attack in Lemma \ref{lemma 2} rises to $100\%$ at the beginning of the attack. This is due to the fact that with $\eta=10^{10}$, such an attack has an extremely large amplitude at initial time. On the contrary, by choosing $\tilde{\eta}=0.01$, the amplitude of the attack in Theorem \ref{theorem 3} increases exponentially over time, and its alarm rate becomes $100\%$ after $200$ time steps. Besides, we consider that the attacker estimates the coding matrix and redesigns the false data as $a_{ij}^*(k)=\tilde{M}_{ij}^{-1}(k)a_{ij}(k)$. By comparing Figs. \ref{simulation 3} and \ref{simulation 4}, we can see that without fully cracking ${M}_{ij}$, the attacker can only postpone the time of being detected.

\begin{figure}[t]
\begin{minipage}[t]{1\linewidth}
\centering
\includegraphics[width=1\linewidth]{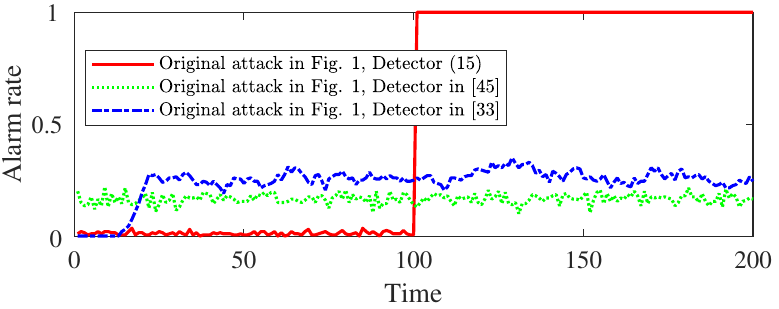}
 \caption{Comparison between the detector (\ref{detector1}), the stochastic protector in \cite{yang2019distributed}, and the relative entropy based detector in \cite{mustafa2022secure} with respect to detecting the attack in Fig. \ref{simulation 1}.}
 \label{simulation 51}
\end{minipage}
\end{figure}

It should be noted that for the distributed estimator (\ref{DKF}), recent advances have proposed some attack detection technologies such as the stochastic protector in \cite{yang2019distributed} and the relative entropy based detector in \cite{mustafa2022secure}. To be specific, the former employs a time-varying threshold as the benchmark for detecting $z_{ij}(k)$, while the latter adopts the K-L divergency between $z_{ij}(k)$ and $z_{i}(k)$ to determine whether $z_{ij}(k)$ is attacked. In view of this, we further compare the proposed detector (\ref{detector1}) with the above two in terms of detecting the attack in Fig. \ref{simulation 1}. In Fig. \ref{simulation 51}, we assume that the system is normal during $[0,100]$, while the attack occurs during the remaining time period. As can be seen, the proposed detector (\ref{detector1}) can detect the attack in Fig. \ref{simulation 1} with the probability of $100\%$, and maintain a low false alarm rate in the absence of the attack. On the contrary, both the detectors in \cite{yang2019distributed} and \cite{mustafa2022secure} are unable to significantly identify the attack, since their detection rates almost close to their false alarm rates. The primine reason is that those detectors rely on the statistical properties of $z_{i}(k)$ and $z_{ij}(k)$, which remain the same before and after the attack in Fig. \ref{simulation 1}. It implies that the insecurity condition in Theorem \ref{theorem 1} is not limited to the $\chi^2$ detector (\ref{detector}), but can be further extended to the other detectors based on $z_{i}(k)$ and $z_{ij}(k)$. In a word, the proposed detector is superior in detecting the attack in Fig. \ref{simulation 1} compared with those in \cite{yang2019distributed} and \cite{mustafa2022secure}.

\section{Conclusion}\label{Section6}
In this work, we study the insecurity of distributed consensus filtering under resource-constrained attackers, who can intrude a subset of channels to maintain strict stealthiness of two types of residues in each node, and diverge the estimation error of distributed sensor networks to infinity. From the perspective of the attacker, we derive the necessary and sufficient condition to determine whether the distributed consensus filtering has the aforementioned security vulnerabilities. Accordingly, from the perspective of the defender, we propose two protection strategies against attacks, which are based on the Euclidean distance among local estimates and the coding scheme, respectively. It is proven that the former can reduce security risks of distributed estimation to a certain extent, while the latter can fully compensate for the security loophole when the coding matrix is not cracked by the adversary. Moreover, we prove that the combined usage of the above two protection strategies helps to relax the feasibility condition required for the coding matrix. Finally, to balance the trade-off between security and coding costs, we provide a procedure for selecting security-critical channels for encoding. Notice that in the field of distributed consensus estimation, there are various types of filtering algorithms apart from the one (\ref{DKF}). Besides, the strict stealthiness of the attack in this paper is a special case of the $\epsilon$-stealthiness in existing literature. Furthermore, adversaries can launch attacks not only on the channels, but also on the sensors. Given these, our future research directions include analyzing the security vulnerabilities of different distributed estimation algorithm with time-varying consensus gain under the $\epsilon$-stealthy attack, and extending existing security countermeasures such as the moving target defense to protect the security of distributed estimation under sensor attacks.

\begin{ack}
We thank Shanghai Institute for Mathematics and Interdisciplinary Sciences (SIMIS) for their financial support. The authors are grateful for the resources and facilities provided by SIMIS, which were essential for the completion of this work.
\end{ack}

\appendix
\section*{Appendix}
\section{Proof of Lemma \ref{lemma 1}}\label{appendix2}
We first consider that the first condition is not satisfied. Based on the definition of $\rho_1^i(k)$, (\ref{appendix1_equation1}) can be rewritten as $\mathbb{E}[\Vert\Delta\hat{x}_i^a(k)\Vert_2]\leq 2\Vert C_i^T(C_iC_i^T)^{-1}\Vert_2\sqrt{\mathrm{tr}(\Sigma_{i})}$ if $\mathrm{rank}(C_i)=n$. Then, we consider that the second condition is not satisfied. Since $\mathbb{E}[\Delta z_i^a(k)]=\mathbb{E}[\Delta z_{ij}^a(k)]=0$, we have $\mathbb{E}[\tilde{\rho}^i(k)]=0$. Thus, by taking the expectation on the left and right sides of (\ref{appendix1_equation2}), we have $\Xi^i \alpha^i(k+1)=A\Xi^i\tilde{\alpha}^i(k)$. Since $\Xi^ix=A\Xi^iy$ does not have any nontrivial solution of $x$, it implies that $\alpha^i(k+1)=0$. Hence, for $\mathrm{rank}(C_i)=m_i<n$, we can derive that $\mathbb{E}[\Vert\Delta\hat{x}_i^a(k+1)\Vert_2]=\mathbb{E}[\Vert\rho_3^i(k+1)\Vert_2]\leq 2\Vert C_i^T(C_iC_i^T)^{-1}\Vert_2\sqrt{\mathrm{tr}(\Sigma_{i})}$. Similarly, when $\mathrm{rank}(C_i)<m_i$ and $\mathrm{rank}(C_i)<n$, its upper bound has the same form as the one of $\mathrm{rank}(C_i)=n<m_i$. The proof is thus completed.

\section{Proof of Theorem \ref{theorem 2}}\label{appendix3}
We prove the necessity at first. Similarly, $\Delta\hat{x}_i^a(k)$ can be rewritten as $\Delta\hat{x}_i^a(k)=\tilde{\Xi}^i \alpha^i(k)+\rho^i(k)$. Note that $\Delta\hat{x}_j^a(k)+a_{ij}(k)=\Xi^i \alpha^{ij}(k)+\rho^{ij}(k)$. Then, based on  (\ref{DKF_attack_difference}), we have
\begin{align}
\tilde{\Xi}^i \alpha^i(k+1)
                          &=A\tilde{\Xi}^i \alpha^i(k)-\varepsilon A\sum\limits_{j\in \mathcal{N}_i}[\tilde{\Xi}^i \alpha^i(k)-(1-\gamma_{ij})\tilde{\Xi}^j \nonumber\\
                           & \alpha^j(k)]+\varepsilon A\sum\limits_{j\in \mathcal{N}_i}[\gamma_{ij}\Xi^i \alpha^{ij}(k)]+\tilde{\rho}^i(k),  \label{appendix3_equation1}
\end{align}
where $\tilde{\rho}^i(k)$ is similar to the one in (\ref{appendix1_equation2}). Define $\alpha(k)=[(\alpha^1(k))^T,...,(\alpha^N(k))^T]^T$, and $\check{\alpha}(k)=[...,(\check{\alpha}^i(k))^T,...]^T$, where $\check{\alpha}^i(k)=[...,(\alpha^{ij}(k))^T,...]^T$. Based on (\ref{appendix3_equation1}), one has
\begin{align}
\mathrm{diag}(\tilde{\Xi}^i) \alpha(k+1)&=[I_N-\varepsilon(L+A_{\gamma})]\otimes A \mathrm{diag}(\tilde{\Xi}^i) \alpha(k)+  \nonumber\\
                                &\varepsilon \mathrm{diag}(A_{\gamma}^{[i]}\otimes (A\Xi^i))\check{\alpha}(k)+\tilde{\rho}(k),\label{appendix3_equation2}
\end{align}
where $\tilde{\rho}(k)$ is a vector whose $i$th element is $\tilde{\rho}^i(k)$. Hence, $\mathbb{E}[\Vert\tilde{\rho}(k)\Vert_2]$ are also bounded, and do not affect the infinite component of $\Delta\hat{x}_i^a(k+1)$. According to (\ref{appendix3_equation2}), the attack can diverge $\Delta\hat{x}^a(k)$ only if $\mathrm{diag}(\tilde{\Xi}^i)\neq 0$, which means that $\exists i \in\mathcal{V}, \mathrm{rank}(\tilde{C}_i^a)<n$. Besides, from the expectation of (\ref{appendix3_equation2}), the attack should satisfy (\ref{theorem2_equation1}) to keep stealthy. The proof of sufficiency is similar to Theorem \ref{theorem 1}, and thus is omitted here.

\section{Proof of Lemma \ref{lemma 4}}\label{appendix6}
Recall that when the channel $(i,j)$ is attacked, i.e., $\gamma_{ij}=1$, the original attack signal $a_{ij}(k)$ satisfies $\Delta\hat{x}_j^a(k)+a_{ij}(k)=\Xi^i \alpha^{ij}(k)+\rho^{ij}(k)$, where $\Delta\hat{x}_j^a(k)$ is constrained by $\Delta\hat{x}_j^a(k)=\Xi^j \alpha^{j}(k)+\rho^{j}(k)$. Hence, the difference between $z_{ij}(k)$ and the residue induced by $\hat{x}_{ij}^e(k)$ is rewritten as
\begin{align}
\Delta &z_{ij}^a(k)=C_i[\hat{x}_{ij}^e(k)-\hat{x}_j(k)]=C_i[\Delta\hat{x}_j^a(k)+M_{ij}a_{ij}(k)]\nonumber\\
                  &=C_i[M_{ij}-I_n](\Xi^i \alpha^{ij}(k)-\Xi^j \alpha^{j}(k))+\tilde{\rho}^{ij}(k),\label{appendix6_equation1}
\end{align}
where $\tilde{\rho}^{ij}(k)\triangleq C_i[M_{ij}-I_n](\rho^{ij}(k)-\rho^{j}(k))+C_i\rho^{ij}(k)$ and $\mathbb{E}[\Vert\tilde{\rho}^{ij}(k)\Vert_2]$ is bounded as well. If the coding matrix satisfies (\ref{lemma 4_equation1}), it means that when $\Xi^i \alpha^{ij}(k)-\Xi^j \alpha^{j}(k)\neq 0$, the first term of (\ref{appendix6_equation1}) must be a nonzero vector, i.e., $[M_{ij}-I_n](\Xi^i \alpha^{ij}(k)-\Xi^j \alpha^{j}(k))\neq \Xi^i y_1$, where $\forall y_i\in \mathbb{R}^{l_i}$. Notice that $\Vert a_{ij}(k)\Vert_2\leq \Vert\Xi^i \alpha^{ij}(k)-\Xi^j \alpha^{j}(k)\Vert_2+\Vert\rho^{ij}(k)-\rho^{j}(k)\Vert_2$, which indicates that when $\Vert a_{ij}(k)\Vert_2\to\infty$, $\Vert\Xi^i \alpha^{ij}(k)-\Xi^j \alpha^{j}(k)\Vert_2\to\infty$. It yields that $\Vert\Delta z_{ij}^a(k)\Vert_2\to\infty$, which completes the proof.

\section{Proof of Theorem \ref{theorem 5}}\label{appendix7}
By intercepting the data $\vartheta_{ij}(k)$ and $\hat{x}_j^a(k)$, the estimated coding matrix is constructed to satisfy the constraint (\ref{encoder}) as well, i.e., $\vartheta_{ij}(k)=\tilde{M}_{ij}^{-1}(k)\hat{x}_j^a(k)$. Then, one has  $(I_n-M_{ij}(k)\tilde{M}_{ij}^{-1}(k))\hat{x}_j^a(k)=0$, where $\hat{x}_j^a(k)=\hat{x}_j(k)+\Delta\hat{x}_j^a(k)$. For the attacker, the unknown matrix $M_{ij}(k)$ satisfying the above equation has infinite solutions besides $M_{ij}(k)=\tilde{M}_{ij}(k)$. Similar to (\ref{appendix6_equation1}), when the encoded channel $(i,j)$ is attacked, the residue generated by the $i$th sensor is
\begin{align*}
\Delta z_{ij}^a(k)=&C_i[\Delta\hat{x}_j^a(k)+M_{ij}(k)\tilde{M}_{ij}^{-1}(k)a_{ij}(k)]\nonumber\\
=&C_i(M_{ij}(k)\tilde{M}_{ij}^{-1}(k)-I_n)\hat{x}_j(k)\nonumber\\
&+C_iM_{ij}(k)\tilde{M}_{ij}^{-1}(k)(\Xi^i \alpha^{ij}(k)+\rho^{ij}(k)).
\end{align*}
Hence, for the attacker, the mean of $\Delta z_{ij}^a(k)$ depends on $C_iM_{ij}(k)\tilde{M}_{ij}^{-1}(k)\Xi^i \alpha^{ij}(k)$ and its covariance is related to $\hat{x}_j(k)$. Recall that to remain strictly stealthy, the attacker should satisfy $\mathbb{E}[\Delta z_{ij}^a(k)]=0$ and $\mathbb{E}[\Vert \rho^{ij}(k)\Vert_2]\leq\beta^{ij}$. Thus, if $\tilde{M}_{ij}(k)\neq M_{ij}(k)$, the attacker cannot ensure that $\Xi^i \alpha^{ij}(k)$ preserve the zero-mean property of $\Delta z_{ij}^a(k)$. Besides, $\hat{x}_j(k)$ can even diverge the covariance of $\Delta z_{ij}^a(k)$ when $A$ is unstable. In a word, the attack can ensure its stealthiness only if $M_{ij}(k)$ is cracked. However, the probability to select $\tilde{M}_{ij}(k)=M_{ij}(k)$ from the infinite solutions is almost equal to zero. The proof is thus completed.

\bibliographystyle{plain}
\bibliography{autosam}           

\end{document}